\documentclass[fleqn,usenatbib]{mnras}

\usepackage[T1]{fontenc}
\usepackage{ae,aecompl}
 
\usepackage{xcolor}
\usepackage{url}
\usepackage{footnote}

\usepackage{graphicx}	% Including figure files
\usepackage{amsmath}	% Advanced maths commands
\usepackage{tablefootnote}

\usepackage{amssymb}	% Extra maths symbols
\usepackage{siunitx}
\usepackage{pdflscape}

\usepackage{enumitem}
\setlist{leftmargin=7mm}
\title[GRB Follow-ups With GOTO]{Searching For Fermi GRB Optical Counterparts With The Prototype Gravitational-Wave Optical Transient Observer (GOTO)}

\author[Mong et al.]{Y. -L. Mong,$^{1,2}$\thanks{E-mail: yik.mong@monash.edu}
K. Ackley,$^{1,2,3}$
D. K. Galloway$^{1,2}$,
M. Dyer$^{4}$,
R. Cutter$^{3}$,
\newauthor
M. J. I. Brown$^{1}$,
J. Lyman$^{3}$,
K. Ulaczyk$^{3}$,
D. Steeghs$^{3}$,
V. Dhillon$^{4}$,
\newauthor
P. O'Brien$^{5}$,
G. Ramsay$^{6}$,
K. Noysena$^{7}$,
R. Kotak$^{8}$,
R. Breton$^{9}$,
\newauthor
L. Nuttall$^{10}$,
E. Pall\'{e}$^{11}$,
D. Pollacco$^{3}$,
E. Thrane$^{1,2}$,
S. Awiphan$^{7}$,
\newauthor
U. Burhanudin$^{4}$,
P. Chote$^{3}$,
A. Chrimes$^{3}$,
E. Daw$^{4}$,
C. Duffy$^{6}$,
\newauthor
R. Eyles-Ferris$^{5}$,
B. Gompertz$^{3}$,
T. Heikkil\"{a}$^{8}$,
P. Irawati$^{7}$,
M. Kennedy$^{9}$,
\newauthor
T. Killestein$^{3}$,
A. Levan$^{3}$,
S. Littlefair$^{4}$,
L. Makrygianni$^{4}$,
T. Marsh$^{3}$,
\newauthor
D. Mata-Sanchez$^{9}$,
S. Mattila$^{8}$,
J. Maund$^{4}$,
J. McCormac$^{3}$,
D. Mkrtichian$^{7}$,
\newauthor
J. Mullaney$^{4}$,
E. Rol$^{1,2}$,
U. Sawangwit$^{7}$,
E. Stanway$^{3}$,
R. Starling$^{5}$,
\newauthor
P. Str{\o}m$^{3}$,
S. Tooke$^{5}$,
K. Wiersema$^{3}$
\\ \\
$^{1}$School of Physics \& Astronomy, Monash University, Clayton VIC 3800, Australia\\
$^{2}$OzGrav: The ARC Centre of Excellence for Gravitational Wave Discovery, Clayton VIC 3800, Australia\\
$^{3}$Department of Physics, University of Warwick, Coventry, West Midlands, CV4 7AL, UK\\
$^{4}$Department of Physics and Astronomy, University of Sheffield, Sheffield, S3 7RH, UK\\
$^{5}$School of Physics and Astronomy, University of Leicester, University Road, Leicester, LE1 7RH, UK\\
$^{6}$Armagh Observatory \& Planetarium, College Hill, Armagh, BT61 9DB, Co.Armagh, Northern Ireland\\
$^{7}$National Astronomical Research Institute of Thailand,  260  Moo 4, T. Donkaew,  A. Maerim, Chiangmai, 50180, Thailand\\
$^{8}$Department of Physics and Astronomy, University of Turku, FI-20014 Turun yliopisto, Finland\\
$^{9}$Department of Physics and Astronomy, University of Manchester, M13 9PL, UK\\
$^{10}$Institute of Cosmology and Gravitation, University of Portsmouth, Dennis Sciama Building, Burnaby Road, Portsmouth, PO1 3FX, UK\\
$^{11}$Instituto de Astrof\'{i}sica de Canarias, E-38205 La Laguna, Tenerife, Spain
}

\date{Accepted XXX. Received YYY; in original form ZZZ}

\pubyear{2021}

\begin{document}
\newcommand{\Msun}{$M_{\odot}$}
\newcommand{\Lsun}{$L_{\odot}$}
\newcommand{\Rsun}{$R_{\odot}$}
\newcommand{\solar}{${\odot}$}
\newcommand{\comment}[1]{\textcolor{red}{{\texttt{#1}} }}

\label{firstpage}
\pagerange{\pageref{firstpage}--\pageref{lastpage}}
\maketitle

\begin{abstract}
The typical detection rate of $\sim1$ gamma-ray burst (GRB) per day by the \emph{Fermi} Gamma-ray Burst Monitor (GBM) provides a valuable opportunity to further our understanding of GRB physics. However, the large uncertainty of the \emph{Fermi} localization typically prevents rapid identification of multi-wavelength counterparts. We report the follow-up of 93 \emph{Fermi} GRBs with the Gravitational-wave Optical Transient Observer (GOTO) prototype on La Palma. We selected 53 events (based on favourable observing conditions) for detailed analysis, and to demonstrate our strategy of searching for  optical counterparts. We apply a filtering process consisting of both automated and manual steps to 60\,085 candidates initially, rejecting all but 29, arising from 15 events. With $\approx3$ GRB afterglows expected to be detectable with GOTO from our sample, most of the candidates are unlikely to be related to the GRBs. Since we did not have multiple observations for those candidates, we cannot confidently confirm the association between the transients and the GRBs. Our results show that GOTO can effectively search for GRB optical counterparts thanks to its large field of view of $\approx40$ square degrees and its depth of $\approx20$ mag. We also detail several methods to improve our overall performance for future follow-up programs of \emph{Fermi} GRBs.
\end{abstract}


%%%%%%%%%%%%%%%%%%%%%%%%%%%%%%%%%%%%%%%%%%%%%%%%%%

%%%%%%%%%%%%%%%%% BODY OF PAPER %%%%%%%%%%%%%%%%%%
\section{Introduction}\label{sec:introduction}
It is generally believed that $\gamma$-ray bursts (GRBs) come from core-collapse supernovae \citep[SNe;][]{mw99,mwh01,wb06,w0011} or compact binary coalescence \citep[CBC;][]{elp89,pac91,kp93,bbm13,rpn13}. They are divided into two classes called long and short GRBs based on the duration of the $\gamma$-ray burst with the boundary historically set at $\sim2\,{\rm s}$ \citep{kmf93,sbb08,sbb11,zsy12,qll13}. Although GRBs have been studied for more than 50 years since the first discovery on 2 July 1967 \citep{kso73}, it was the simultaneous detection of GW170817 and GRB170817A \citep{lv17,aaa17,aaa17b,gvb17,sfk17} that the progenitor origin of short GRBs as CBC sources was confirmed. 

The Burst and Transient Source Experiment \citep[BATSE;][]{gpm13} launched in 1991, established the first step in GRB research by showing the cosmological origin of GRBs. In order to explain the isotropic equivalent luminosity of GRBs, which could be up to $\sim10^{54}\,{\rm ergs}$ \citep{fks01}, the fireball model was introduced \citep{pac86,goo86,sp90,rm92,psn93,wrm97,vie97,mrr02}. This model suggests that due to relativistic outflows, the $\gamma$-ray emission is highly beamed with the beaming angle of $\theta_b\sim1/\Gamma$, where $\Gamma$ is the bulk Lorentz factor of the outflow \citep{rho99,sph99}. The effect of relativistic beaming can significantly reduce the energy such that it can explain the energy scale of the GRB. The fireball model also invokes internal and external shocks to explain the prompt $\gamma$-ray emission and the afterglow \citep{bm76,rm94,bel00}.

In 1997, BeppoSAX \citep{bbp97} first enabled multi-wavelength observations of GRB afterglows \citep{cfh97,vgg97,fkn97}. The Wide Field Cameras (WFCs) on board BeppoSAX can track GRB Monitor (GRBM) triggers to locate the GRBs with a precision of $\approx3$ arcmin \citep{jmb97}. The detection of X-ray counterparts can help identify the optical counterpart, which can then be used to identify the host galaxy and constrain the luminosity of the GRBs with its redshift \citep{mdk97}. BeppoSAX also helped GRB980425 to be identified with SN1998bw \citep{gvv98}, which suggested that the origin of long GRBs is core-collapse SNe \citep{mw99,mwh01}.

Today the discovery of GRBs is mainly made through the \emph{Swift} and \emph{Fermi} satellites. There are two narrow-field instruments on board \emph{Swift} --- the X-Ray Telescope \citep[XRT;][]{bhn05} and the Ultra-Violet and Optical Telescope \citep[UVOT;][]{rkm05} --- which can search for the electromagnetic counterparts to the \emph{Swift} Burst Alert Telescope \citep[BAT;][]{bbc05} triggers.  However, the \emph{Fermi} Gamma-ray Burst Monitor \citep[GBM;][]{mlb09,kmp20} dominates the GRB detection rate, finding $\sim1$ per day, but with much larger location uncertainties compared to  \emph{Swift}. Other than \emph{Fermi}-GBM, \emph{Fermi} also equipped the Large Area Telescope \citep[LAT;][]{aaa09} in order to detect high-energy gamma-ray emission from the GRB sources. The detection rate of \emph{Fermi}-LAT is $\sim10$ per year \citep{kz15}. As \emph{Fermi} does not have either an X-ray or optical telescope on-board, and for GRBs for which there is no joint \emph{Swift} detection, it is important to follow-up the \emph{Fermi} GRB detections in order to expand the sample of GRB afterglows available for GRB studies.

In general, there are two observing strategies that are used to find GRB optical afterglows: serendipitous and target-of-opportunity (ToO) follow-ups. For instance, \citet{ack21} demonstrates how to identify serendipitous fast transients, such as optical afterglow and kilonova, detected by \emph{Zwicky Transient Facility} (ZTF) using {\tt ZTFReST}, an automated filtering and follow-up infrastructure. The greatest science return of using this strategy is the proof of the existence of the orphan afterglow \citep{cup15,caa20,hpb20}. However, it is not an efficient way to searching for optical GRB afterglows as it is a blind search of the sky \citep{npg02,huh20}. On the other hand, ToO follow-up directly searches for the optical counterparts of the detected GRB by tiling skymaps generated by the \emph{Fermi} or \emph{Swift} instruments \citep{sck13,skc15,cac19,asa21}. However, the sizes of these skymaps can vary substantially, from square degrees to arcseconds, depending on the gamma-ray instrument. Targeting potential host galaxies  and multi-color observations are usually used in ToO follow-ups, which help to improve search efficiency and identify the origin of the candidates. Other than ZTF, there are other observatories which actively follow-up GRBs, such as Global MASTER-Net \citep{lkg12}, Nordic Optical Telescope \citep[NOT;][]{da10} and GROWTH-India Telescope (GIT), DDOTI \citep{wlt16}.

The Gravitational-wave Optical Transient Observer (GOTO) is an optical robotic telescope that consisted of four 40\,cm f/2.5 unit telescopes (UTs) in its prototype design (GOTO-4) until 1 November 2019 and was upgraded to a full eight UTs (GOTO-8) in November 2019 \citep[][Steeghs et. al in prep]{dsg20}. The primary objective of the GOTO project is to detect the optical counterpart of  gravitational-wave (GW) events detected by the LIGO-Virgo collaboration (LVC) network. GOTO followed up LVC triggers distributed during the third observing period, O3 (between April 2019 and March 2020), and the details of GOTO's O3 follow-up strategies and results are presented in \citet{gcs20}. In the absence of active GW triggers, GOTO also listens to automated alerts for GRBs from satellites such as \emph{Fermi} and \emph{Swift}. In the absence of any GRB or GW triggers, GOTO performs an all-sky survey on a routine basis. The pixel scale of GOTO is $\approx1.2$ arcsec per pixel. GOTO can reach depths of $\sim$21 mag in the broad Baader L filter with a co-added set of $4\times90\,{\rm s}$ observations. Combined with a wide field of view (FoV) of $\approx40\,$square degrees in the GOTO-8 prototype design, GOTO has the capability for searching for \emph{Fermi} GRB optical counterparts.

In this paper we demonstrate how we perform automated follow-up and detail strategies for how we narrow down the number of potential candidates of a \emph{Fermi}-GRB optical afterglow. We also discuss the expected improvement in our follow-up strategy for the current GOTO-8 and for forthcoming upgrades to the network. In \S\ref{sec:obs_strategy}, we describe our current observing strategy and in \S\ref{sec:data_prep} we indicate how we filter the GRB events automatically. The flow of manual filtering, which is different from the automated tasks from \S\ref{sec:data_prep}, in the final stages is outlined in \S\ref{sec:manual_filter}. Our final results are described in \S\ref{sec:results}. Finally, we discuss the planned expansion of the GOTO network and future follow-up strategy in \S\ref{sec:discussion}.

\section{GOTO Observing Strategy for GRBs}\label{sec:obs_strategy}
GOTO is a fully-autonomous telescope and is controlled by the GOTO Telescope Control System \citep[G-TeCS;][]{gtecs18,gtecs20}. The G\nobreakdash-TeCS software includes an alert monitor called the \texttt{sentinel}, which receives GW and GRB alerts released through the NASA Gamma-ray Coordinates Network (GCN) \citep{GCN}. Once follow-up targets have been generated by GOTO \texttt{sentinel} for an event they are added to the GOTO observation database from which the automated scheduler will select targets and instruct the telescope to move the mount to the requested position and take the requested exposures. The entire system is automated and results in fast reaction times on minute timescales; if the sky position is visible and GOTO is able to observe then exposures can begin in $\sim$1-2 minutes of the alert being received. The fast response of GOTO was successfully demonstrated by the follow-up observations of GW events detected in the LIGO-Virgo collaboration (LVC) O3 run \citep{gcs20}. 

When not following-up alerts GOTO carries out an all-sky survey based on a fixed grid of tiles, which builds up an archive of reference images covering all points on the sky. When a GW or GRB alert is received its probability region is mapped onto the same grid as the survey, which allows any new images to be compared to the reference templates taken at the same position (see Section~\ref{sec:data_prep}). Mapping the event skymap onto the survey grid is done using the \texttt{goto-tile} Python package \citep{dyerthesis}. 

For \emph{Fermi} GRB events, a two-dimensional Gaussian skymap is independently created by \texttt{goto-tile}, centered at the location reported by a GCN. Since \emph{Fermi}-GBM shows a systematic error of $\approx3.6^\circ$ \citep[at 68\% confidence level;][]{cbg15}, we use the quadrature sum of the systematic error and the statistical error (which is included in the GCN) as the 68\% confidence region of the skymap. The measurements of the GRB location reported on GCN are only useful for generating our own Gaussian skymap as there are delays of $\sim10$ minutes between receiving the the initial alert and the generation of the official \emph{Fermi} skymap. We will ultimately replace the artificial Gaussian skymap with the official \emph{Fermi} skymap once it has been received (see \S\ref{sec:gaussian_against_official} for more details about the difference between our Gaussian skymap and the \emph{Fermi} skymap).

Once the skymap has been created by \texttt{goto-tile}, the contained probability within each GOTO tile is calculated. For this work, a large fraction of our observing period overlapped with LIGO-Virgo O3. In order to avoid observations taking excessive amounts of time away from completing our template set before O3 and the follow-up observations of GW events during O3, only the five tiles with the highest probability are added into the observation database and only if they each include at least 5\% of the overall localization probability error region.

As well as determining which tiles to target for follow-up observations, the G\nobreakdash-TeCS \texttt{sentinel} also defines the number of times each tile should be observed and the delay between observations. In order to confirm whether an observed transient source is associated with the GRB at least two observations are required within 1 day after the trigger, to confirm the fast-decay nature of the GRB optical afterglow \citep{kkz10, kkz11}. Therefore, for GRB observations the \texttt{sentinel} uses the strategy \texttt{TWO\_FIRST\_ONE\_SECOND}, which requires two observations to be taken within the first day after the trigger followed by another single exposure the day after. It should be noted that G\nobreakdash-TeCS uses a ``just-in-time'' scheduling system, which while it is flexible enough to allow for very fast initial observations, it cannot guarantee the exact timings for subsequent observations. Even accounting for delays due to poor weather or if the Sun rises before a second observation can be taken, it is possible that a higher-priority target can be added into the scheduler queue, taking precedence over the GRB target while both targets are visible.

Finally, when targets are added into the observations database, the exposure sets are then similarly defined. For \textit{Fermi}-GRB sources a set of four, 90~second exposures are taken using GOTO's wide \textit{L} filter (400-700~nm). In this work, all GOTO magnitudes have been calibrated in the standard way as described in \citet{dsg20}.

This work is an independent project of GOTO designed to test approaches for GRB afterglow searches, which does not represent the GOTO standard filtering procedures as described in \citet{dsg20}.

\section{Data Preparation}\label{sec:data_prep}
\subsection{Data Selection}
In this paper, we outline the results of the GRB follow-up from 22 February 2019 to 7 June 2020. We used the GOTO-4 prototype configuration until 1 November 2019. Beyond that date, we used the upgraded to GOTO-8 prototype configuration.

\begin{figure}
    \centering
    \includegraphics[width=\columnwidth]{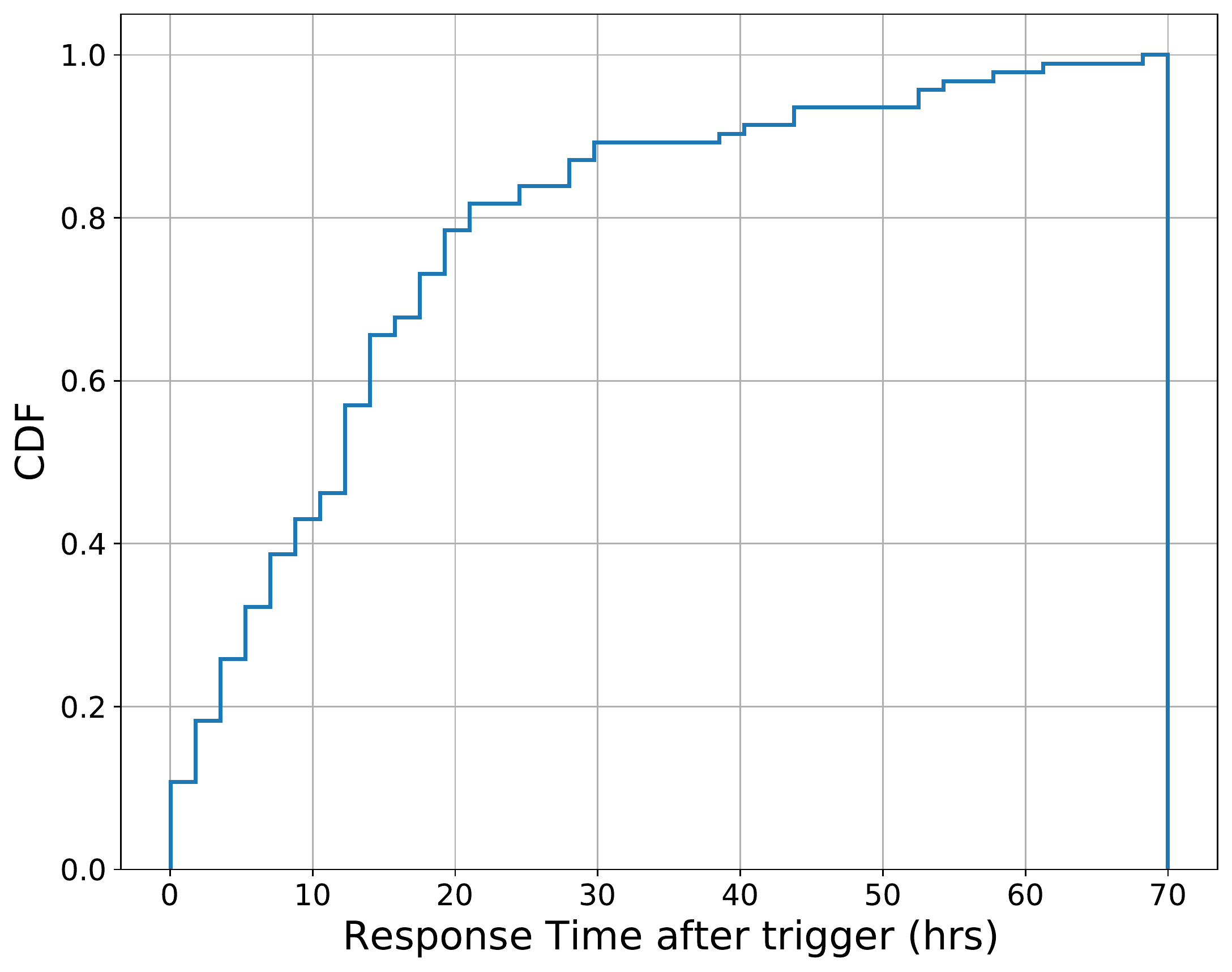}
    \caption{Cumulative distribution of GOTO response time for all 92 observed \emph{Fermi}-GRB triggers. The response time has the range between 141 seconds and 70 hours. The median response time is 12.9 hours.}
    \label{fig:goto_response}
\end{figure}

From 22 February 2019 to 7 June 2020, there were 508 \emph{Fermi}-GBM triggers reported on \emph{Fermi} trigger GCN. Among those, there were 376 triggers located at declination greater than $-30$ degrees. Due to further restrictions of the GRB locations and timing and weather conditions at the site, GOTO observed a total of 93 of these alerts. Since one of the alerts (TriggerNum: 596808579) was 
later classified as a non-GRB trigger, we further reduce the observed targets to 92. The GOTO response times range between $\approx140$ seconds to $\approx3$ days, with the median response time $\approx13$ hours (see Figure \ref{fig:goto_response}). There are three reasons which can greatly delay the GRB follow-up observations. Firstly, bad weather or poor site conditions are the most common issues preventing us from observing the event immediately after the trigger. Secondly, the event location is not always observable when triggered. Finally, since GW event triggers always have a higher follow-up priority, we do not interrupt any ongoing follow-up observation of the GW event due to the GRB trigger. Therefore, we postpone the GRB follow-up until the GW follow-up has finished.

Motivated by the nature of expected GRB afterglow decay timescales \citep{kkz10}, we focus only on events for which observing commenced within 16 hours after the trigger. There are 58 out of 92 events fulfilling this criterion. The objective of this work is to perform a systematic search for optical counterparts to GRBs which were detected by solely by \emph{Fermi}-GBM. It also helps to inform our future GW search strategies. Therefore, we filter out 5 events which have joint-detection with \emph{Swift}-BAT such that we have 53 \emph{Fermi} events in total. The details of all events followed up with GOTO are shown in Table \ref{tab:events}. It includes the event time and the location reported in GCN of each GRB. We also include the response time of our first follow-up observation after the trigger. The coverage percentages shown in Table \ref{tab:events} are calculated based on the official skymap provided by \emph{Fermi}. The expected probability of detecting the optical afterglow for each GRB estimated base on Eq. \ref{eq:prob_oa} is also included in Table \ref{tab:events}.

\subsection{Image Processing}
The raw images are processed with the GOTO standard pipeline \citep[][]{dsg20}, which includes bias correction, dark-frame subtraction and co-addition of individual images. All images used to perform analyses in this paper are co-added median science images.

A stacked template image of four individual exposures is used to perform image subtraction. It acts as a reference image taken before the GRB trigger. The set of templates are updated regularly by tiling the sky on a fixed grid, as described in Section \ref{sec:obs_strategy}. We align the template to match with the science image by using Python package \texttt{spalipy}\footnote{\url{https://github.com/Lyalpha/spalipy}}. We use \texttt{hotpants} \citep{bec15} to perform image subtraction. Since not all fields have templates taken before the science images, we use the last observation prior to the GRB trigger as our reference image if the template does not exist. In the image subtraction step, we generate 1229 difference images in total for all 53 \emph{Fermi}-GRB events.

\subsection{Filtering Processes}
After the difference imaging process, we gather all source detections in the difference images recovered by \texttt{SExtractor} \citep{ba96}. Prior to any filtering, there are 15\,049\,101 detections in total among the 1229 difference images. For those GRBs with $\approx100\%$ coverage, the set of images can cover more than the $1\sigma$ region of the skymap. However, since the current observing strategy used is covering a fixed number of grids instead of covering certain proportion of the skymap, we account for all the detections in those difference images.

\subsubsection{Real-bogus classification}\label{sec:rb_class}
After the image processing, we apply the real-bogus classification on the difference images to separate the bogus detections and the real detections by using supervised machine learning technique \citep{mag20,kls21}. We choose the decision boundary such that the upper limits of the false positive rate and the missed detection rate are $2\%$ and $4\%$ respectively. After the real-bogus classification, there are 60\,085 detections which are classified as ``real'' on the difference images. All ``real'' detections proceed to the next filtering step.

\subsubsection{Bad data pre-filtering}\label{sec:bad_data_filter}
Any detection that satisfies one of the following conditions is classified as a bad detection or statistical noise;
\begin{enumerate}
    \item the physical position lies within 50 pixels from the edges (the image size of GOTO is $8176\times6132$ pixels with an angular resolution of $\approx1.2$ arcsec per pixel), 
    \item the full-width half-maximum (FWHM) of the detection on the difference image is greater than the $95$th-percentile of the FWHM distribution of the science image detections,
    \item the signal-to-noise ratio $<3\sigma$,
    \item the detection is fainter than the limiting magnitude of the science image,
    \item the detection is brighter than $14\,{\rm mag}$.
\end{enumerate}
Of the 60\,085 detections, 17\,058 are classified as bad data or noise. Those detections are then rejected from our candidate list in order to reduce the time costs of cross-matching and human vetting process. Therefore, we further reduce the number of potential candidates down to 43\,027 by filtering the set of ``real'' detections using image-based characteristics.

\subsubsection{Catalog Cross-matching}\label{sec:cat_xmatch}
Variable objects are commonly classified as real in the real-bogus classification (see \S\ref{sec:rb_class}). To effectively reject the variables and the known stellar objects from our candidates, we use Python package \texttt{catsHTM} \citep{so18} to perform cross-matching with other public domain data of stellar survey catalogs (see Table \ref{tab:catalogs}) including GAIA-DR2 \citep{gbv18}, 2MASS \citep{csv03}, AAVSO \citep{whp17}, PS1 \citep{hlt15}, UCAC4 \citep{cmm16}, IPHAS \citep{bfd14} and SIMBAD \citep{wod00}. 

GRB optical afterglows are often found close ($\lesssim10\,{\rm kpc}$) to their host galaxy \citep{bkd02,fb13,bbf16}. A search for the presence of any known galaxy within 1 arcmin of the detection can help us to prioritize those host-associated candidates. Candidates with multiple nearby galaxies are prioritized first as they have a higher chance of association with the GRB. With a matching radius of $10\,{\rm kpc}$ \citep[95--percentile of the GRB physical offset distribution;][]{bbf16}, 1 arcmin can cover down to $z\sim0.008$, which is similar to the redshift of the second closest GRB, GRB980425 \citep{tsc98}. For those galaxies with known redshift, $z$, we can also estimate the isotropic equivalent energy $E_{\rm iso}$ of the GRBs.

We use SIMBAD \citep{wod00} and the GLADE catalog \citep{dgd18} to perform galaxy cross-matching. SIMBAD provides reliable classification based on the physical nature of the objects. The GLADE catalog is created by combining four galaxy catalogs, GWGC, 2MPZ, 2MASS XSC and HyperLEDA, to achieve a high completeness up to $\sim40\,{\rm Mpc}$ \citep{dgd18}. The completeness of the current galaxy catalogs is one of the main issue throughout GRB follow-up studies. However, it cannot be resolved easily. There are 2\,095 out of 43\,027 candidates found next to a known galaxy within 1 arcmin. We label all those candidates with \texttt{near\_galaxy=1}, otherwise \texttt{near\_galaxy=0}.

\begin{table}
	\centering
	\caption{Catalogs used in cross-matching with our GRB-associated candidates.}\label{tab:catalogs}
	%\label{tab:data_set}
	\begin{tabular}{lcc}
		\hline
		Catalog Name & Criteria & Reference\\
		\hline
		GAIA-DR2 & -- & (1) \\
		2MASS & -- & (2) \\
		AAVSO & -- & (3) \\
		APASS & -- & (4) \\
		\hline
		PS1 & $0\geq\log|\mathcal{L}_{\rm ipsf}|>-3$ & (5) \\
		& ${\tt iPSFMag}<19.5$ &\\
		\hline
		UCAC4 & ${\tt Yale\_FLAG}=0$ & (6) \\
		& ${\tt LEDA\_FLAG}=0$ &\\
		& ${\tt ExtCat\_FLAG}=0$ &\\
		& ${\tt 2MASSExt\_FLAG}=0$ &\\
		\hline
		IPHAS & ${\tt mergedclass}=-1$ & (7) \\
		& ${\tt pstar}>0.9$ & \\
		\hline
		SIMBAD & Non-Galaxy Types & (8) \\
		\hline
		\multicolumn{2}{l}{\footnotesize(1) \citet{gbv18}}\\
		\multicolumn{2}{l}{\footnotesize(2) \citet{csv03}}\\
		\multicolumn{2}{l}{\footnotesize(3) \citet{whp17}}\\
		\multicolumn{2}{l}{\footnotesize(4) \citet{hlt15}}\\
		\multicolumn{2}{l}{\footnotesize(5) \citet{cmm16}}\\
		\multicolumn{2}{l}{\footnotesize(6) \citet{zgh13}}\\
		\multicolumn{2}{l}{\footnotesize(7) \citet{bfd14}}\\
		\multicolumn{2}{l}{\footnotesize(8) \citet{wod00}}
	\end{tabular}
\end{table}

\begin{figure}
    \centering
    \includegraphics[width=\columnwidth]{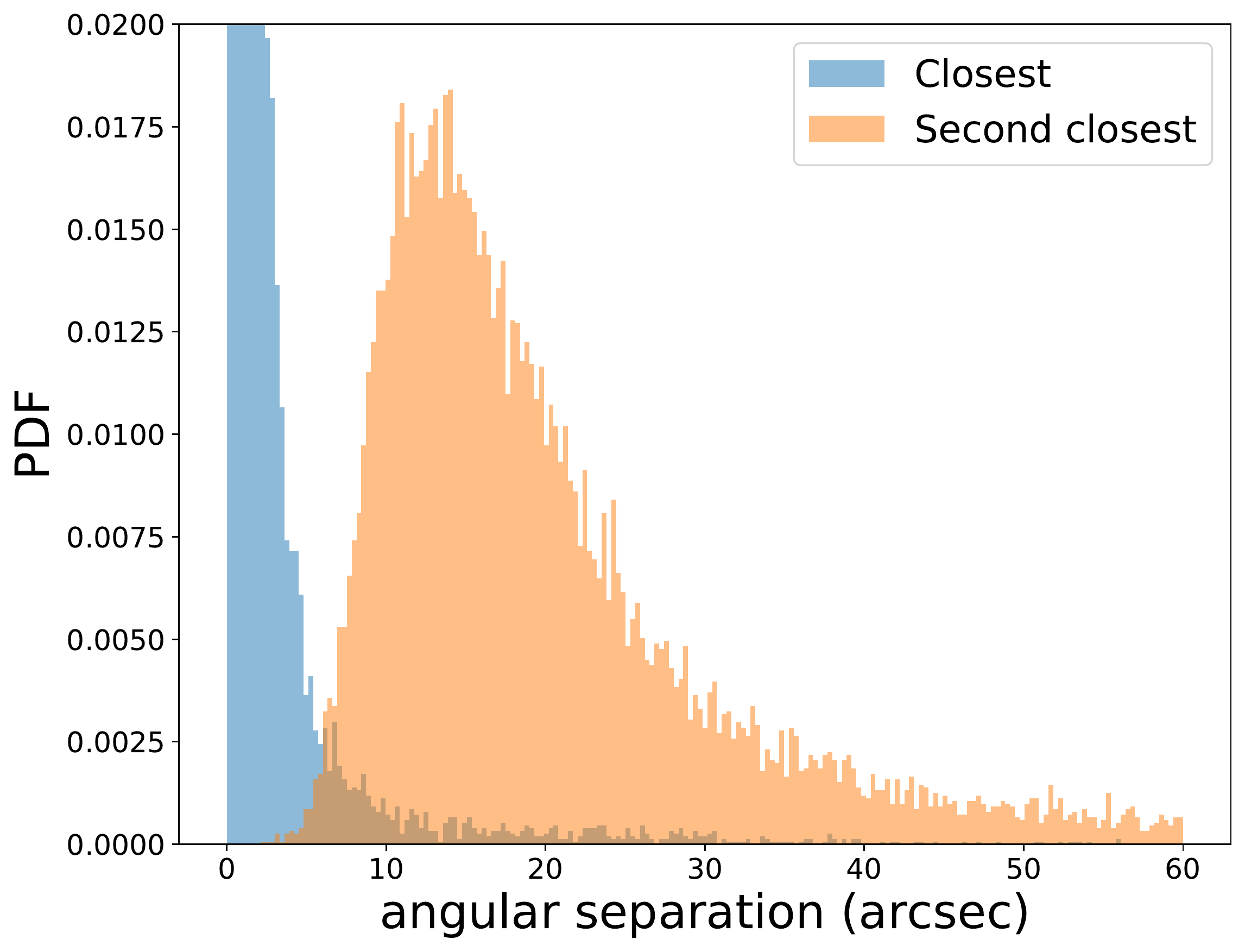}
    \caption{Distribution of the angular separation between the detections on the difference images and the two closest detections on the science images. The {\it blue} histogram shows the distribution of the angular offsets of the closest sources between the science images and the difference images due to the image subtraction. The {\it orange} histogram shows the distribution of the angular separation from the second closest, but unrelated, source.}
    \label{fig:offset}
\end{figure}

To determine whether a real detection on the difference image is associated with a known stellar object from another catalog, we set a decision boundary on the cross-matching offset. The average offset between the real detections on the difference images and the known sources from other catalogs is:
\begin{align}
    \theta=\sqrt{\theta_{\rm ds}^2+\theta_{\rm sc}^2}~,
\end{align}
where $\theta_{\rm ds}$ is the astrometric offset between the difference image and the science image, and $\theta_{\rm sc}$ is the astrometric offset between the science image and the catalog, respectively. Under the assumption that the astrometry from GOTO gives $\theta_{\rm sc}\approx0$, we have $\theta\approx\theta_{\rm ds}$. Figure \ref{fig:offset} shows the distributions of $\theta_{\rm ds}$ and the angular separation between the detections on the difference images and the second closest detections on the science images, which they are presumably not associated with each other. We use the 95-percentile of $\theta_{\rm ds}$, which is at $\approx6$ arcsec, to be our decision boundary for claiming the association between the detection on the difference image and the cross-matched source. However, from the {\it orange} histogram in Figure \ref{fig:offset}, there are 0.7\% real detections having at least one more real detection within $6$ arcsec. Therefore, this filtering step might cause $\approx0.7\%$ false rejection rate.

We label all known stellar sources with \texttt{known\_source=1}, otherwise \texttt{known\_source=0}. We then only reject those candidates with \texttt{known\_source=1} and \texttt{near\_galaxy=0}. This helps us to filter out those known stellar objects which do not associate with known galaxy. With this filtering process, there are 12\,568 candidates remain on our candidate list. Other than that, there could be a possibility that some unresolved galaxies, which are not recorded in SIMBAD and the GLADE, are misclassified as point sources in those stellar catalogs. We cross-match the candidates rejected in this step with NASA/IPAC Extragalactic Database\footnote{The NASA/IPAC Extragalactic Database (NED) is operated by the Jet Propulsion Laboratory, California Institute of Technology, under contract with the National Aeronautics and Space Administration.} (NED) to obtain the misclassification rate as $\approx0.2\%$, which is estimated by the ratio between the number of the cross-matched galaxies in NED and the total number of the known stellar sources.

\subsubsection{Minor planet checking}
We pass our remaining candidates to cross-match with the Minor Planet Catalog (MPC) online\footnote{\url{https://minorplanetcenter.net/}}. Any candidates with a known cross-matched MP within $10$ arcsec at the observing epoch are rejected. There are 10\,126 detections remaining in our candidate list.

\subsubsection{Multi-detection filtering}
To further verify whether a detection is associated with the GRB or not, we cross-match the detection with our source database. Following the same procedure to perform cross-matching in \S\ref{sec:cat_xmatch}, we associate the two cross-matched detections within $6$ arcsec as the same detection.

Any candidates which have been recovered multiple times as real before the GRB trigger or far beyond the triggering time should be excluded as they are not associated with the GRB. Since GRB optical afterglows are fast transients following a power-law decay
\begin{align}
    F\propto t^{-\alpha}~,
\end{align}
where $F$ is the afterglow flux, $t$ is the time after the trigger and $\alpha$ is the power-law index with a typical value of $\sim1.2$
, we expect that they become too faint to be detected in one to two days after the trigger. However, as long GRBs are typically associated with supernovae, they rise long after the optical afterglow has faded and show much longer decay timescales in a few tens of days once the supernova has peaked in its emission \citep{wes99, ksk19}. We define a GRB-associated period as [$t_0$,$t_0 + 100$,days], where $t_0$ is the GRB trigger time. Any candidates which have been recovered as real detections twice outside of the GRB-associated period are rejected. Finally, we have 6\,276 candidates left after all the automatic filtering processes.

\subsection{Summary of automatic filtering}
In the automated filtering processes, we have filtered out about $90\%$ of the real detections. The filtered detections include bad data (see \S\ref{sec:bad_data_filter} for the definitions of bad data), known sources, minor planets and GRB-unrelated sources.

\section{Manual Filtering}\label{sec:manual_filter}
\subsection{Human vetting}
After automatically filtering about $90\%$ of the real detections, we start our vetting process by human inspection. Human vetting consists of three steps, the manual real-bogus classification, the selection of the potential GRB-related objects and the filtering of cosmic rays and bad data, which may have inadvertently passed the automated stages. 

The first step could be subjective. We suggest that different people review each candidate multiple times to minimize subjective bias. In this work, we do not take the issue of misclassification by humans into consideration. 

For the second step, we select only those detections which satisfy either one of the following criteria, there is a nearby galaxy within $10$ arcsec or it does not exist on the template image. Since GRBs commonly show strong association with their hosts, any detection found next to known galaxies could be promising. The first criterion helps us to extract all detections with small angular offset from a known galaxy. For those GRB sources having angular offset $>10$ arcsec, they should be well separated from their host on the image, which should be included with the second criterion. The second criterion also includes any potential hostless GRB sources.

For the final step, we confirm that any candidate is not a cosmic ray by verifying the existence of the detections on all individual images that make up the median co-added science frame. For those detections which fall within the overlapping region on multiple UTs, we also check to verify that the detections are co-located in the overlapping UTs.

Under the human vetting process, there are 116 candidates passing the above three steps. Among them, there are 55 candidates located within 1 arcmin from at least one known galaxy registered in GLADE or SIMBAD.

\subsection{Transient cross-matching}\label{sec:transient_xmatch}
In this section, we describe our final filtering process. We cross-match each potential candidate with the Lasair \citep{swy19} and Transient Name Server\footnote{\url{https://wis-tns.weizmann.ac.il}} (TNS) candidate databases using their API. Any candidates discovered before the GRB trigger epoch should not be associated with the event. Therefore, we filter out all candidates which have been discovered before the trigger. This step is separated from the catalog cross-matching in \S\ref{sec:cat_xmatch} because {\tt catsHTM} does not support the Lasair and the TNS databases. Also, we do not loop our queries through their databases in order to avoid overloading their servers.

In addition to the filtering according to the discovery epoch, we also filter our candidates by their object types if their classification has been reported by Lasair or TNS. The filtering process excludes the object types which are very unlikely to be associated with the GRB optical counterpart, such as variable stellar objects, type-Ia SN, or AGN. However, we have identified 2 candidates, ZTF18aaegvyd\footnote{\url{https://lasair.roe.ac.uk/object/ZTF18aaegvyd/}} and ZTF18acaujfk\footnote{\url{https://lasair.roe.ac.uk/object/ZTF18acaujfk/}}, which show re-brightening within the first day after the trigger and do not have a confident source classifications. We decide to keep those 2 candidates in our candidate list for further photometric analysis. After the transient cross-matching, we obtain a final list of 29 potential candidates.

\begin{table}
	\centering
	\caption{Number of candidates filtered at each step of transient vetting.}\label{tab:flow}
	%\label{tab:data_set}
	\begin{tabular}{lcc}
		\hline
		Filtering step & \multicolumn {2}{c}{Number of remaining candidates} \\
		& Before & After \\
		\hline
		{\bf Automated Filtering} \\
		Real-bogus classification & 15\,049\,101 & 60\,085 \\
		Bad data pre-filtering & 60\,085 & 43\,027 \\
		Catalog cross-matching & 43\,027 & 12\,568 \\ 
		Minor planet filtering & 12\,568 & 10\,126 \\
		Multi-detection filtering & 10\,126 & 6\,276 \\ 
		{\bf Manual Filtering} \\
		Human vetting & 6\,276 & 116 \\
		Transient cross-matching & 119 & 29 \\
		\hline
	\end{tabular}
\end{table}

\section{Results}\label{sec:results}
There are 29 final candidates which passed through all the filtering processes. In this section, we investigate the candidates primarily through photometric measurements derived from forced photometry. 

We also estimate the expected number of GRB optical afterglows which could be detected for all those 53 GRB events.

\subsection{Forced photometry Analysis}
We use the forced photometry service provided by ZTF and ATLAS\footnote{\url{https://fallingstar-data.com/forcedphot/}} to inspect the lightcurves of all potential candidates. Any flux measurement with signal-to-noise ratio (SNR) greater than $5\sigma$ is considered a detection. In order to avoid treating any subtraction residual on the difference image as real detections, we inspect all cutout images for those detections with ${\rm SNR}>5\sigma$.

Among those 29 candidates, there are 2 candidates $\gtrsim10$ arcsec away from known MPs, 4525 Johnbauer and 6384 Kervin. As we are unable to see any detection from both ZTF and ATLAS forced photometry data, we conclude that they are very likely minor planets. We generate the lightcurves for the remaining 27 candidates between 365 days before the trigger and 365 days after their trigger.

\subsubsection{ZTF18aaegvyd/SN2019env}
As mentioned in \S\ref{sec:transient_xmatch}, ZTF18aaegvyd re-brightened in the first day of the GRB triggers (see Figure \ref{fig:rebrighten_known_source}). ZTF18aaegvyd was spectroscopically classified as a Type-II SN \citep{ssy20}, SN2019env\footnote{\url{https://www.wis-tns.org/object/2019env}} in TNS, at $z=0.0235$ by the Spectral Energy Distribution Machine (SEDM) on the Palomar 60-inch telescope (P60). ATLAS reported the first detection of the SN about 9 hours before the GRB was triggered. Since Type-II SN is not a typical origin of GRB \citep{gvv98,mw99,mwh01}, also with the pre-detection epoch which is inconsistent with the typical delay of the GRB-SN, we conclude that SN2019env was not associated with the GRB event, Fermi578679393.

\subsubsection{ZTF18acaujfk}
\begin{figure}
    \centering
    \includegraphics[scale=0.3]{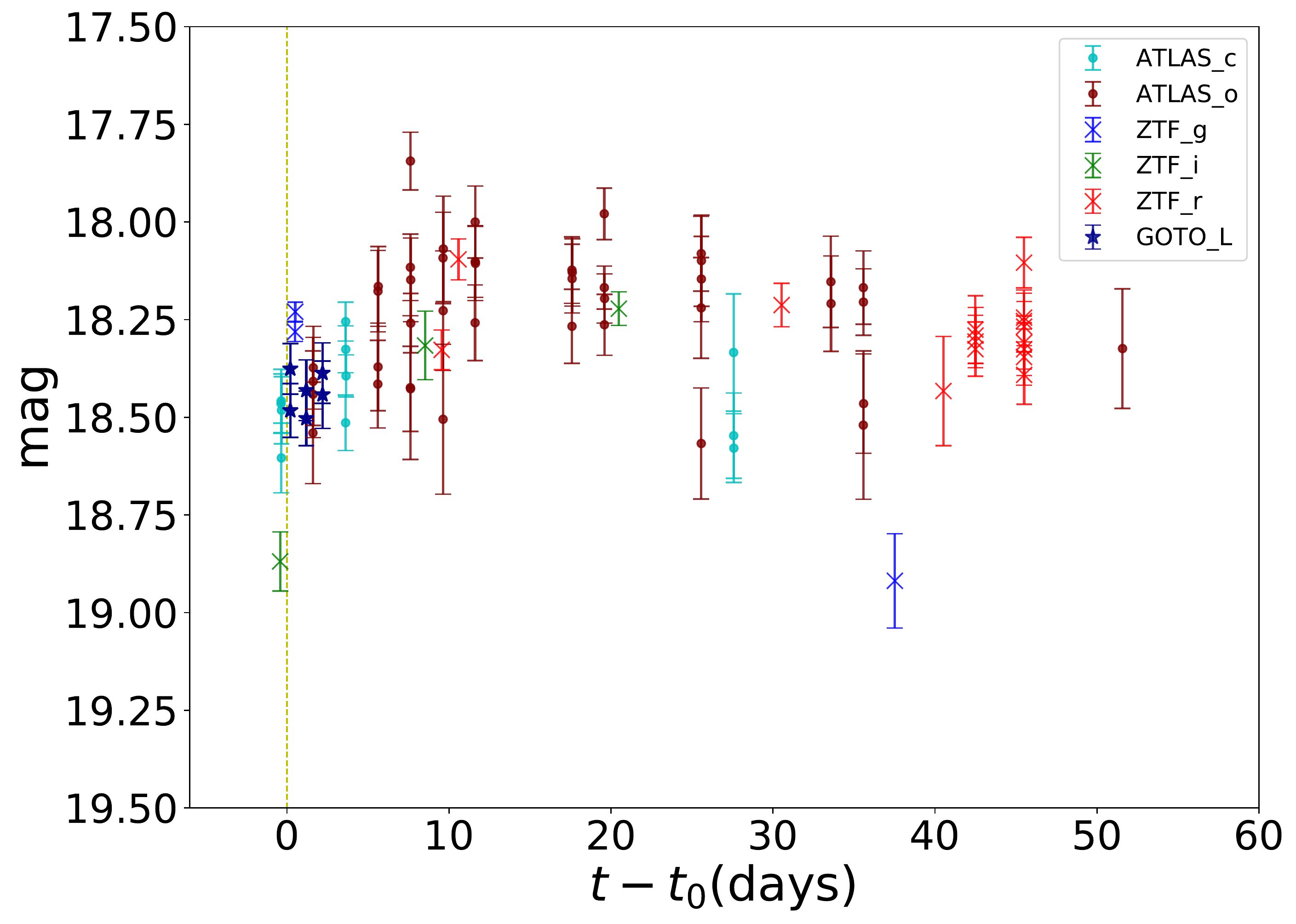}
    \includegraphics[scale=0.3]{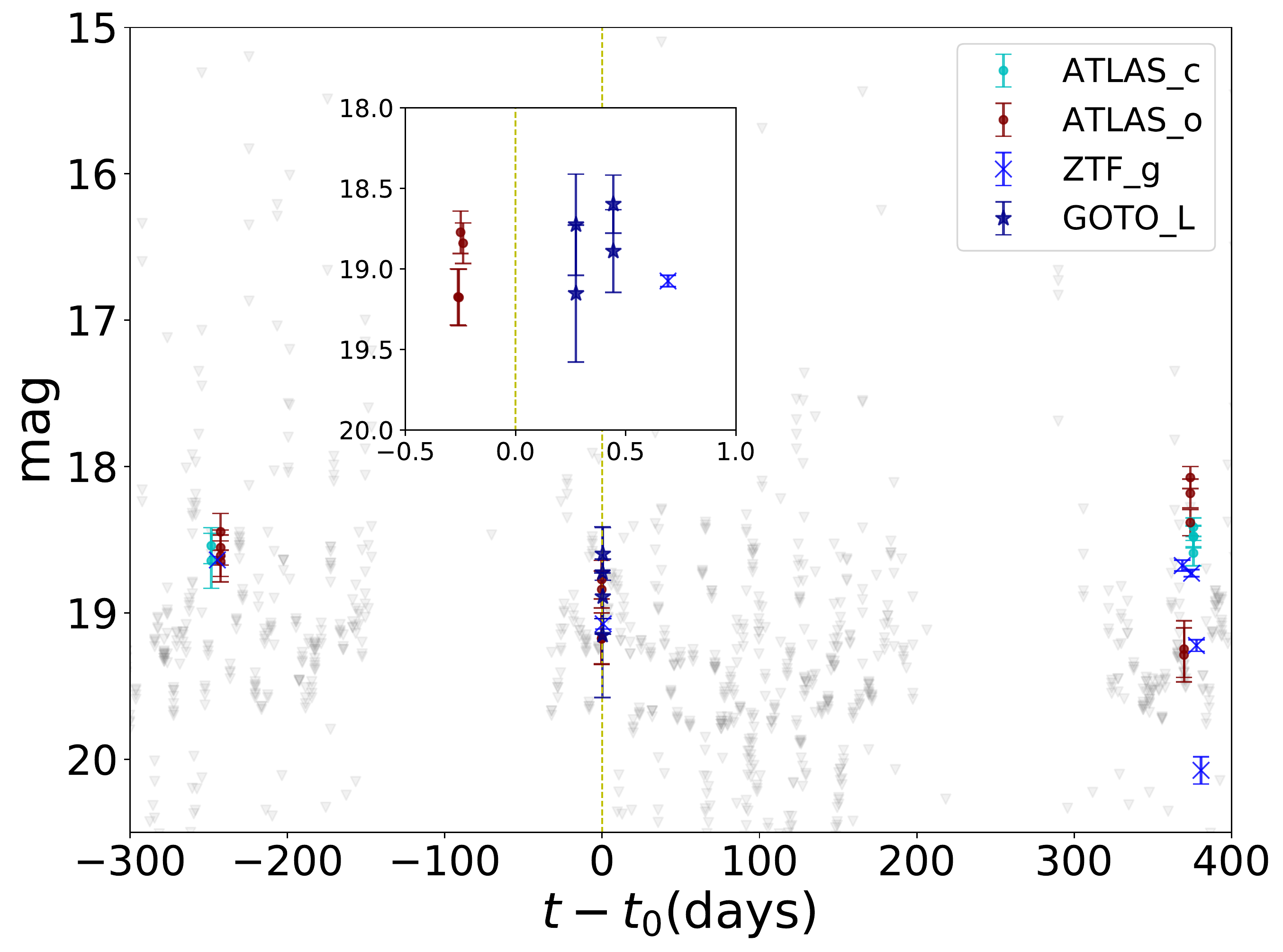}
    \caption{Lightcuves of ZTF18aaegvyd/SN2019env ({\it top}) and ZTF18acaujfk ({\it bottom}). The {\it orange dashed} line indicates the epoch of the GRB trigger. The {\it grey inverted} triangles indicates the limiting magnitudes of ATLAS observations. Both sources show re-brightening on the first day of the GRB trigger.}\label{fig:rebrighten_known_source}
\end{figure}

ZTF18acaujfk is another transient discovered by ZTF which re-brightened on the same day of the GRB trigger. However, the forced photometry data from ATLAS shows that the re-brightening had started within 1 day before the trigger time and the detection only lasted for one day (see Figure \ref{fig:rebrighten_known_source}). The brightness of the source seems to be stable during its active period. Due to the pre-detection from ATLAS, ZTF18acaujfk is unlikely associated with the GRB.

ZTF18acaujfk does not show any detections between 242 days before the trigger and 369 days after the trigger, except for the detection on the day of the trigger. The first detection was claimed by ZTF on 9 October 2018 with $g=18.3$, which was 245 days earlier than the trigger time. However, according to the forced photometry data, it was first detected by ATLAS 33 days before the source was discovered by ZTF. More interestingly, ZTF claimed another SN detection at the same position 1 year after the GRB trigger. Since the typical timescale of SNe is much shorter than a year, if the classification performed by ZTF is correct \citep{ssy20}, we expect that the source would be unrelated to the other events at the same position. However, as this classification has not been spectroscopically confirmed as a SN, we instead interpret this signal as being more consistent with a quasi-periodic variable source.

Cataclysmic variables (CVs) usually show a periodicity of several hours or repeated high amplitude outbursts \citep{rk03, knr16}. Since ZTF18acaujfk was detected three times in around 2 year, it could potentially be CV. For the galactic stellar flare, ZTF18acaujfk matches with the typical active period of a few hours. However, it should consistently be detected. We cannot confirm the transient type of ZTF18acaujfk solely by the the available photometry due to its large variation of the decay timescale from hours to days. In order to further verify the transient type of ZTF18acaujfk, we need to obtain more data by further observations and through spectroscopic classification.

\subsubsection{Other GRB-unrelated transients}
Among the remaining 25 candidates, 3 of them, which are labelled as known sources from catalog cross-matching (see \S\ref{sec:cat_xmatch}), are identified as variable stellar objects from the characteristic variability of their lightcurves. They initially passed our checks and are considered as candidates due to at least one known galaxy within 1 arcmin around the source. 

\begin{figure}
    \centering
    \includegraphics[scale=0.3]{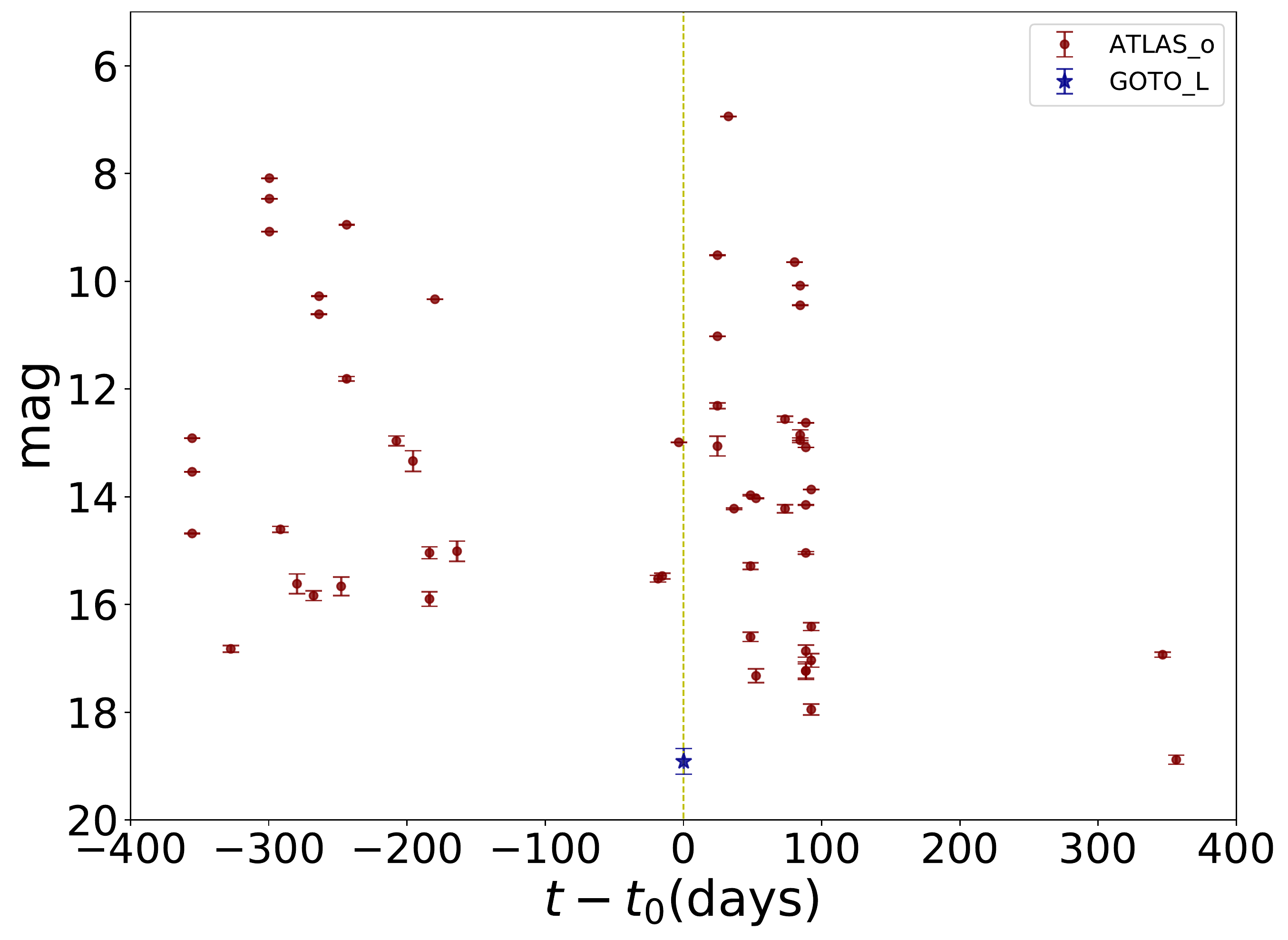}
    \includegraphics[scale=0.3]{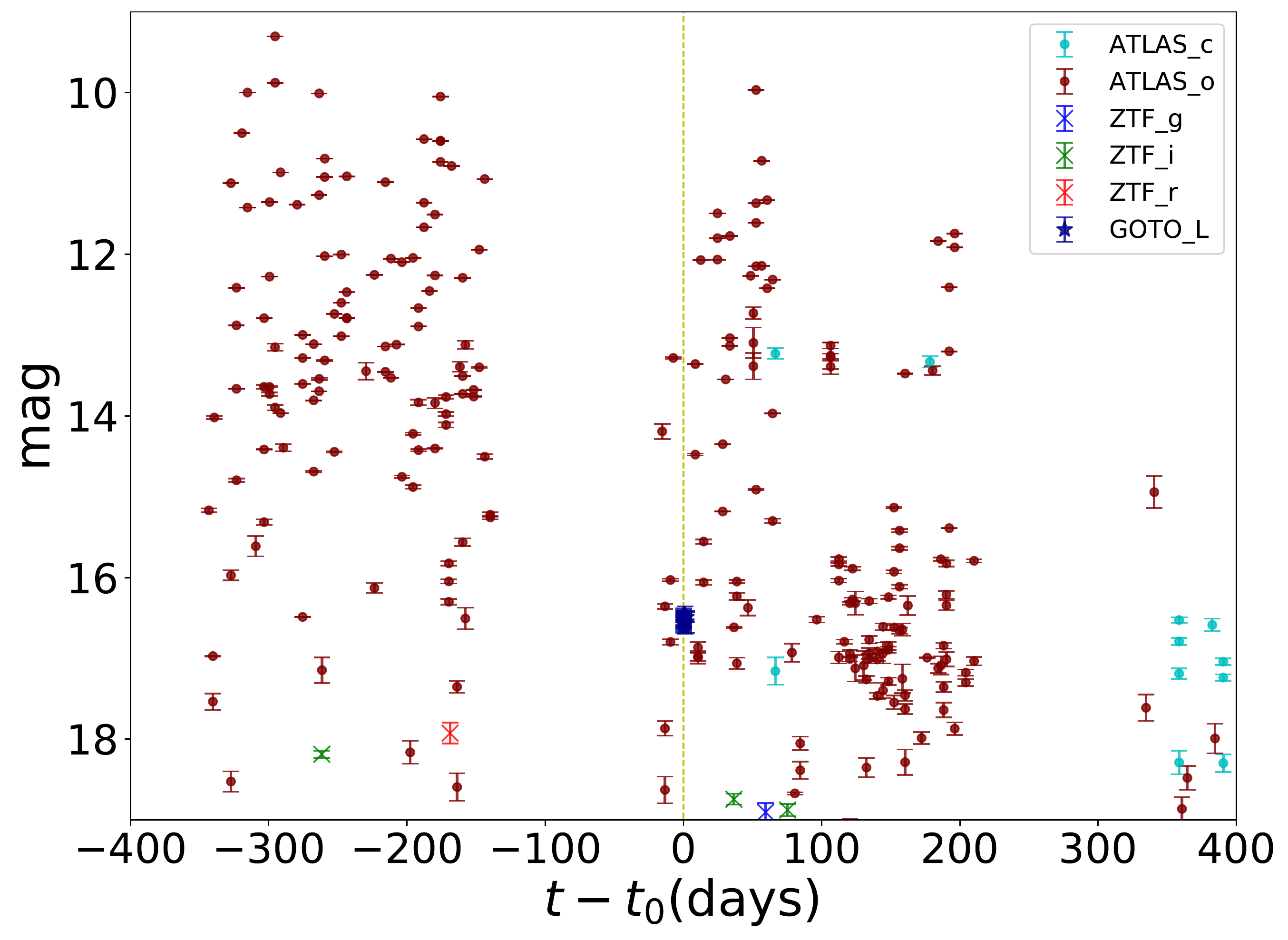}
    \includegraphics[scale=0.3]{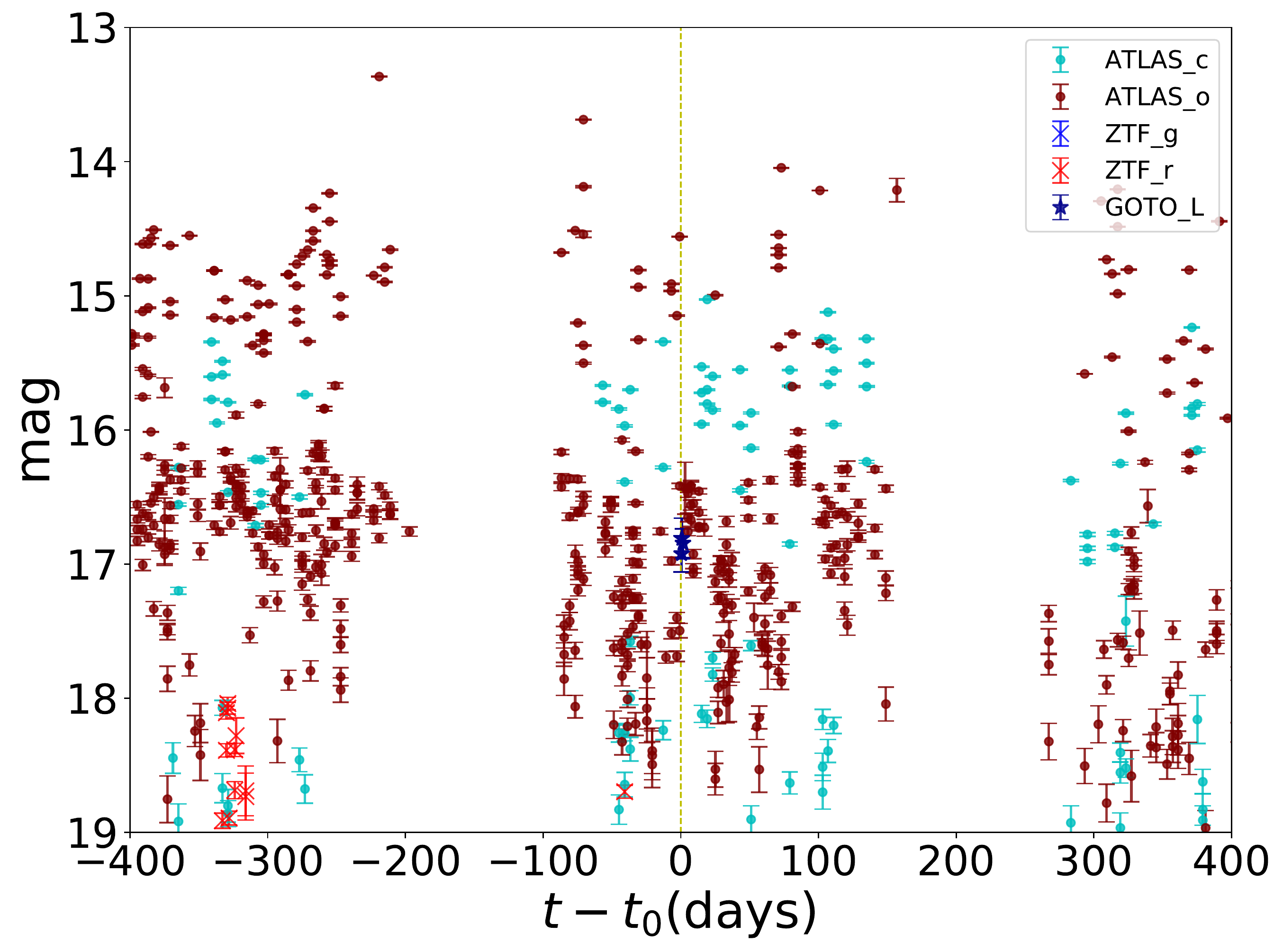}
    \caption{Lightcurves of the candidates which are identified as variable stars.}\label{fig:vs_candidates}
\end{figure}

\begin{figure}
    \centering
    \includegraphics[scale=0.3]{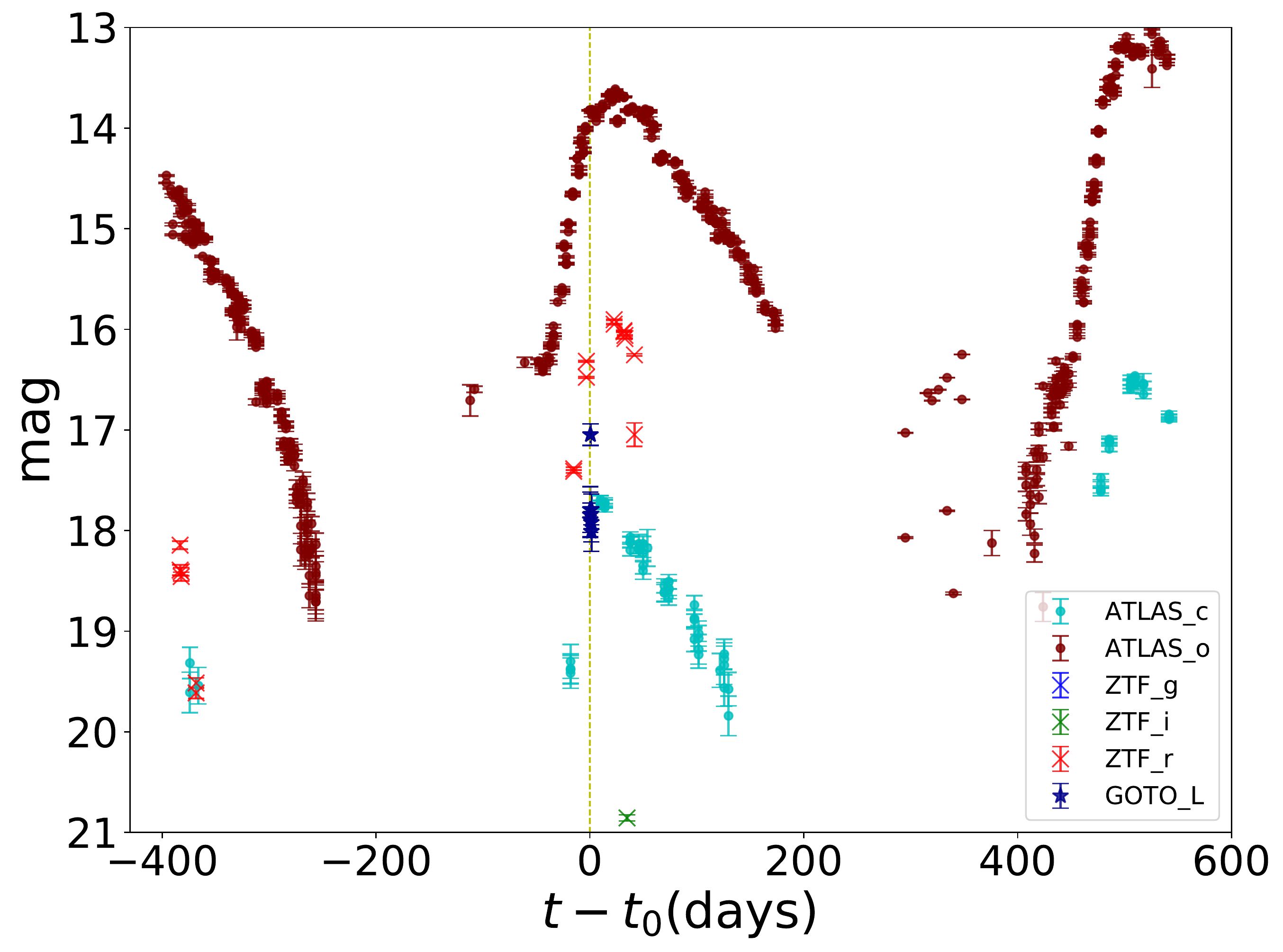}
    \includegraphics[scale=0.3]{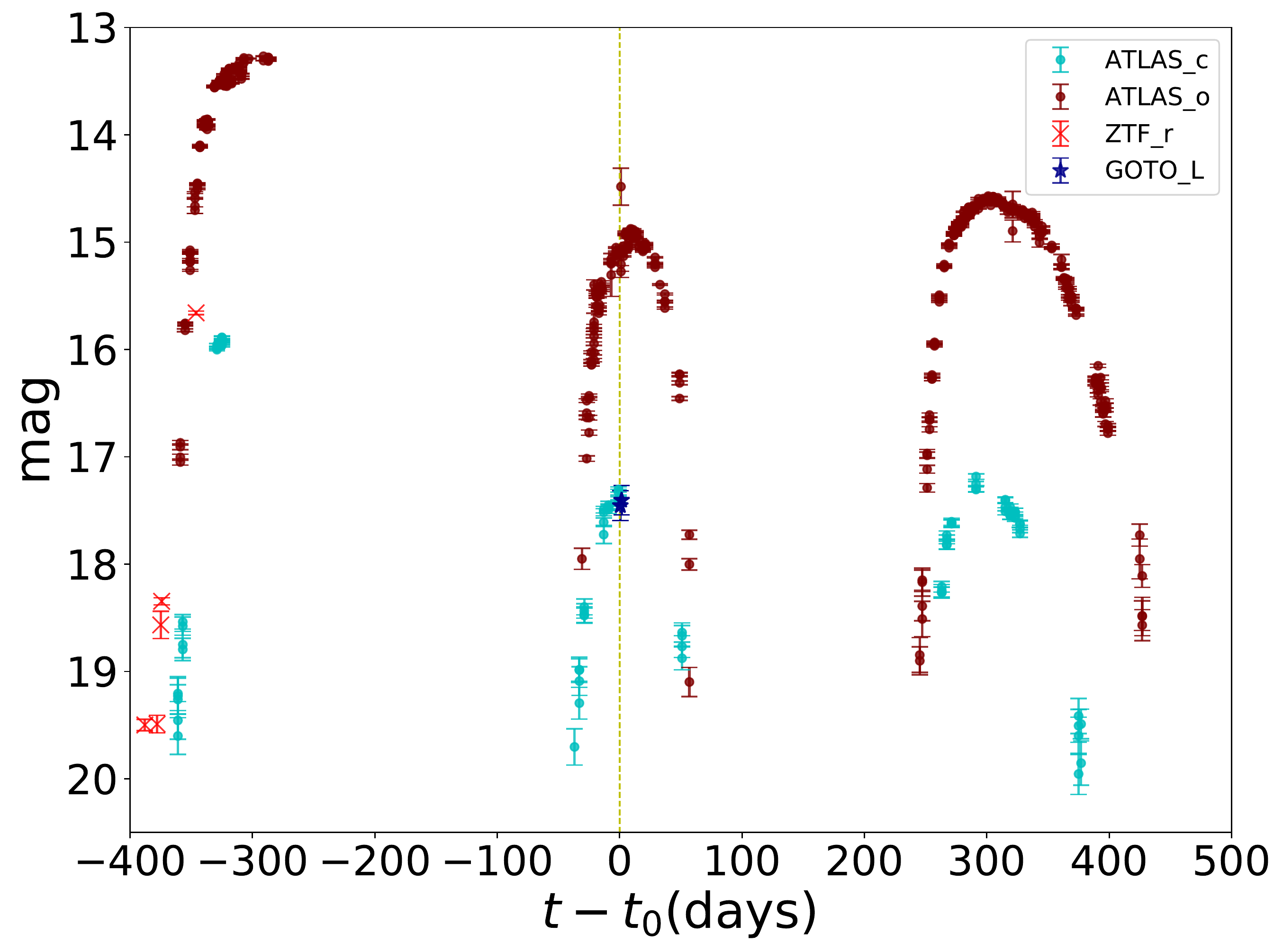}
    \caption{Lightcurves of two unregistered periodic transients. From their periodicities and color, they are very likely to be Mira variables.}\label{fig:mira}
\end{figure}

We have also identified 2 unregistered variable sources. However, they show periodicities of a few hundreds of days with a smooth variability which do not look like typical variable stellar objects (Figure \ref{fig:mira}). In addition, ATLAS shows that they are $\sim$2--4 magnitudes brighter in $o$-filter detection than that in $c$-filter detection. It indicates that they are red in colour, which match with the observational properties of Mira variables.

\begin{figure}
    \centering
    \includegraphics[scale=0.3]{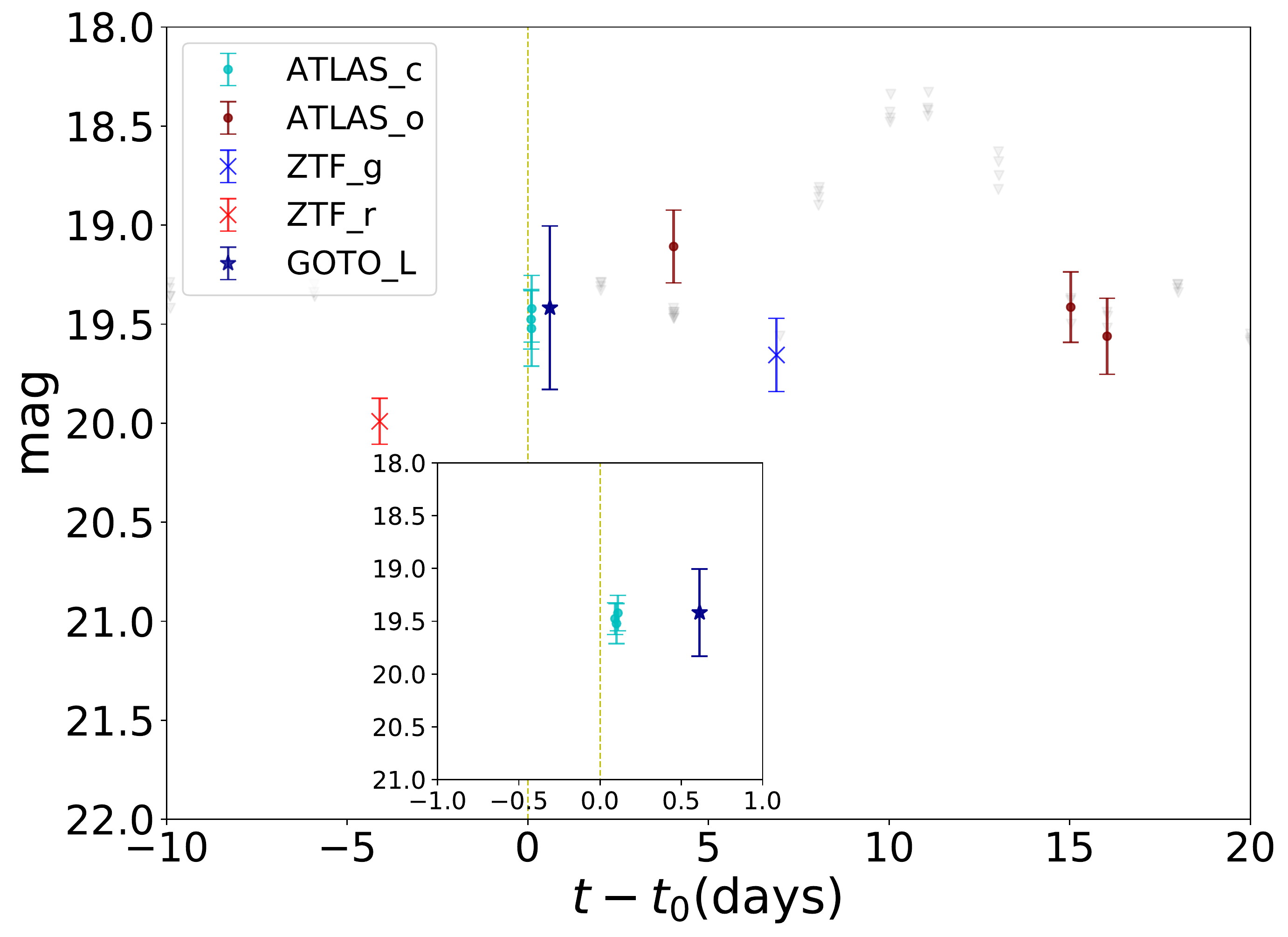}
    \includegraphics[scale=0.3]{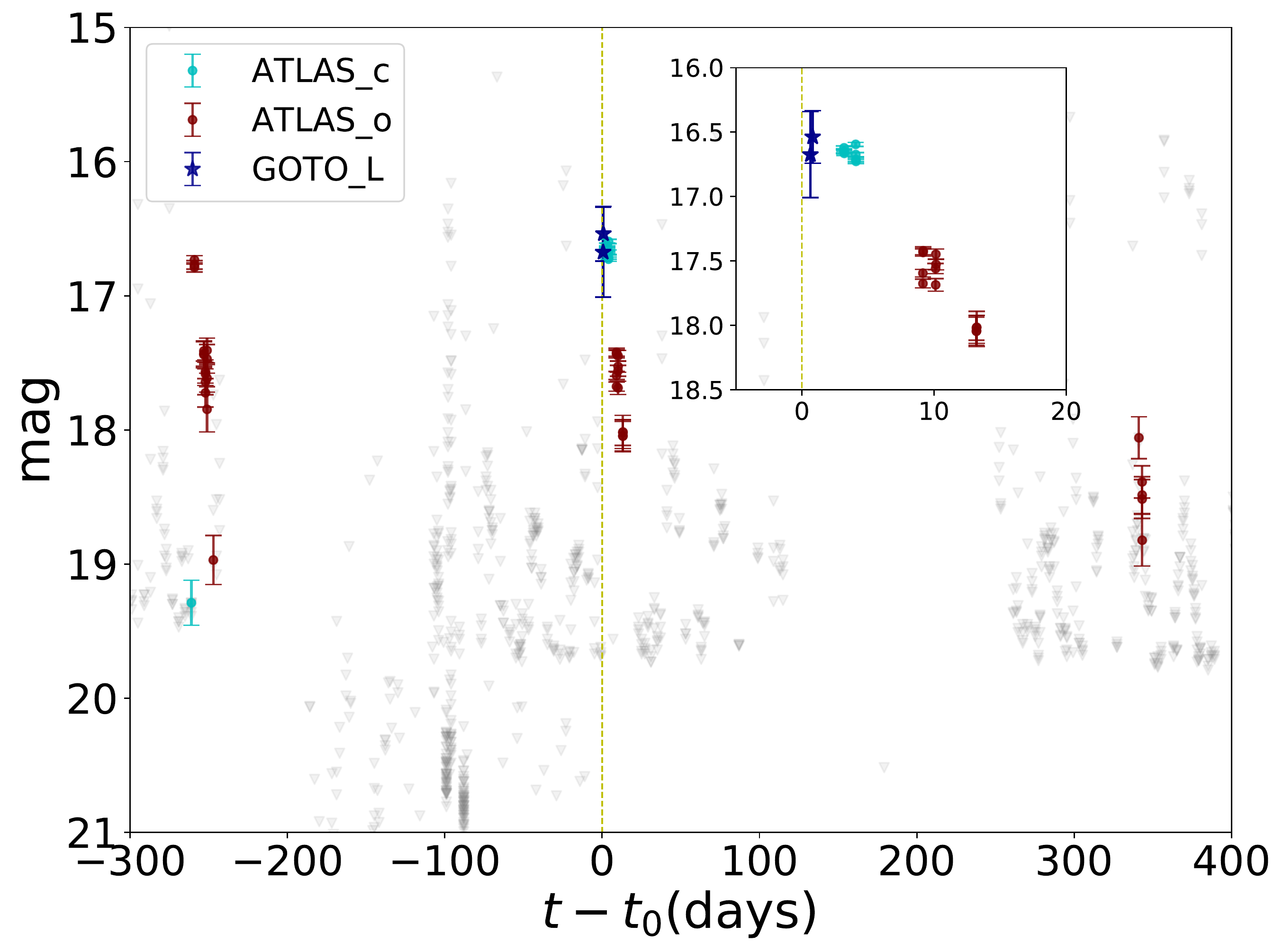}
    \caption{Lightcurves of GOTO103202.04-120240.18 {\it (top)} and GOTO062359.85-185857.69 {\it (bottom)}. The {\it orange dashed} line indicates the epoch of the GRB trigger. The {\it grey inverted} triangles indicates the limiting magnitudes of ATLAS observations. Both of the sources were detected before the GRB trigger times, which imply that they are unlikely to be related to the GRB event.}\label{fig:goto_unknown}
\end{figure}

We found another 2 unknown transients, which are named as GOTO103202.04-120240.18 and GOTO062359.85-185857.69, which showed pre-detections before the triggers by ZTF and ATLAS, respectively (see Figure \ref{fig:goto_unknown}). 

GOTO103202.04-120240.18 was detected by all three instruments, GOTO, ZTF and ATLAS. However, since the first detection obtained by ZTF was around 5 days before the GRB trigger, we conclude that GOTO103202.04-120240.18 was not related to the GRB. From the lightcurve generated using 2-year forced photometry data, the source only stayed active for $\approx20$ days around the GRB trigger. Also, it showed a relatively stable brightness over the entire observing period except for the first ZTF $r$-band detection. We can also see a faint extended source located at the position of GOTO103202.04-120240.18 on Pan-STARRS images. Due to the gap between the 6th and the 7th detections, and without any spectroscopic classification, we are unable to confirm the nature of GOTO103202.04-120240.18.

GOTO062359.85-185857.69 was detected by ATLAS starting $\sim250$ days before the GRB trigger. We do not find any point-source appearing outside the active periods of the source as observed by ATLAS on both ZTF and GOTO images. However, we find a co-located source within archival Pan-STARRS images at the same position of GOTO062359.85-185857.69. The source can be seen in $g$, $r$, $i$ and $z$ bandpasses except $y$-band, which indicates the blue color of the source. Due to the multiple detections from ATLAS and the existence of the stellar source on Pan-STARRS images, it is unlikely to be associated with the GRB. Nevertheless, limiting magnitudes of ATLAS observations in Figure \ref{fig:goto_unknown} show that the source was being monitored around the period of the GRB trigger, which indicates that the re-activation of the source and the GRB triggering epoch are temporally coincident.

\subsubsection{Other potential candidates}
For the remaining 18 candidates. we cannot draw any solid conclusion for them due to the lack of observational data. However, we can still divide them into 3 groups according to similarities in their characteristics.

There are 6 candidates that fall into group 1: GOTO105852.27+491055.26, GOTO175424.66+694237.24, GOTO175722.6+692716.7, GOTO175823.15+694250.58, GOTO180217.26+695300.86 and GOTO190520.52+630533.28. All of these candidates are  within $2$ arcsec from the center of a galaxy. They were not detected by either ZTF or ATLAS. As they are found with low spatial offsets from the galaxies, the resulting sources on their subtracted images do not show clear point-like structures. Therefore, at least another observation within the first day after the trigger can help us to confirm the detection and constrain the decay properties of the afterglow.

There are 7 candidates classified as group 2: GOTO064003.86+294644.32, GOTO051809.33+222414.61, GOTO190511.9-260027.86, GOTO190642.62-211323.51, GOTO191337.34-244759.56, GOTO192152.79-193955.67 and GOTO190259.23-243739.33. Both ZTF and ATLAS have never shown any historical photometric detections of these 7 candidates. Only a single detection or multiple detections taken at the same time were obtained by GOTO for each of them. Therefore, we cannot strongly constrain their lightcurves, including rise time and decay rate. However, we can see a faint point-like object near each of the candidate positions on the Pan-STARRS images. Given that they are stellar objects, these candidates could be M-dwarf flares or GRB optical afterglows which are located right next to a stellar object.

\begin{figure}
    \centering
    \includegraphics[width=\columnwidth]{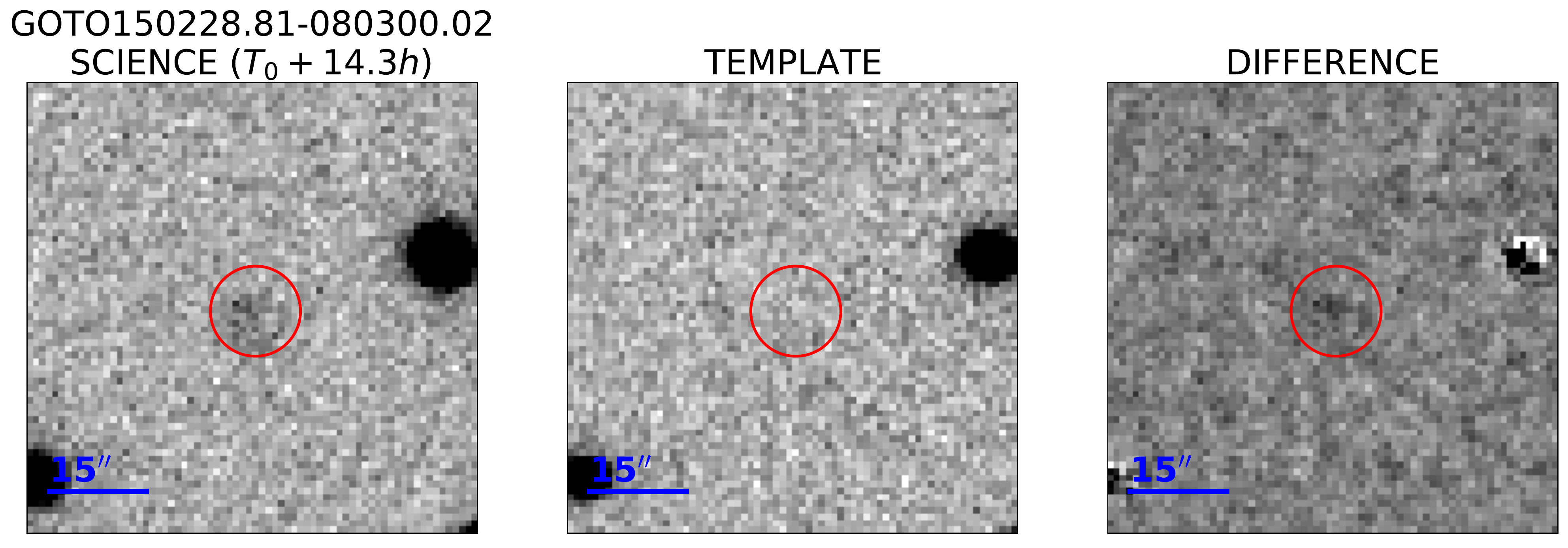}
    \includegraphics[width=\columnwidth]{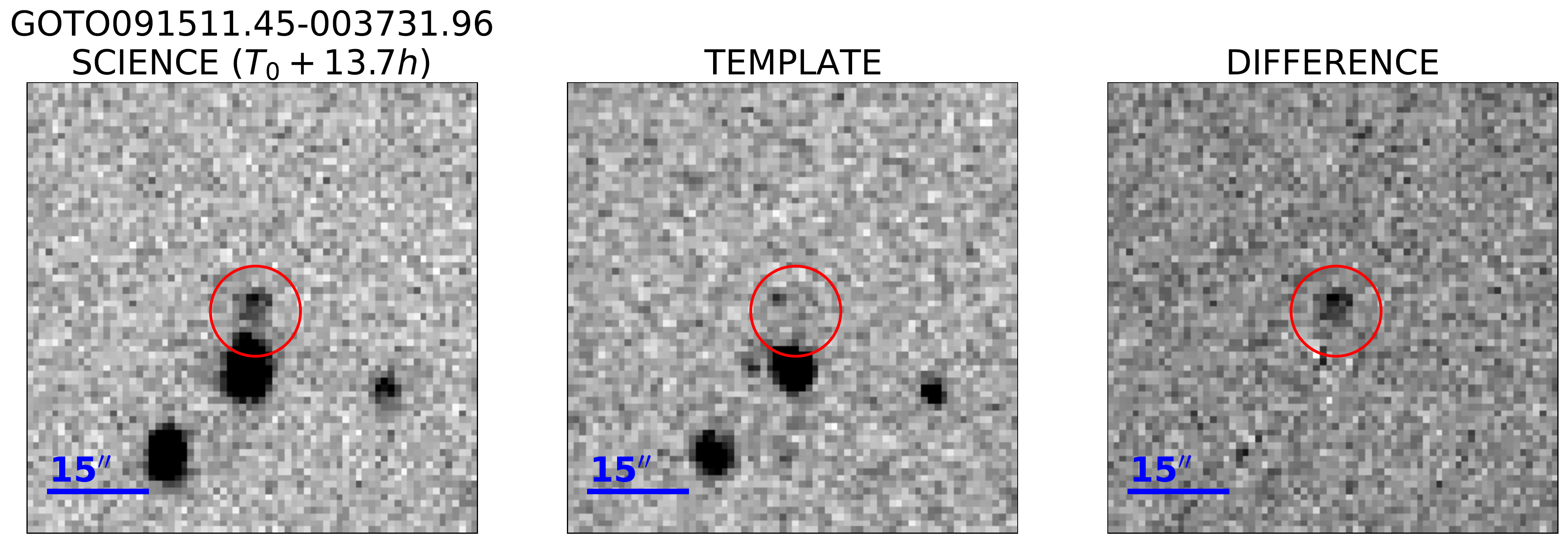}
    \includegraphics[width=\columnwidth]{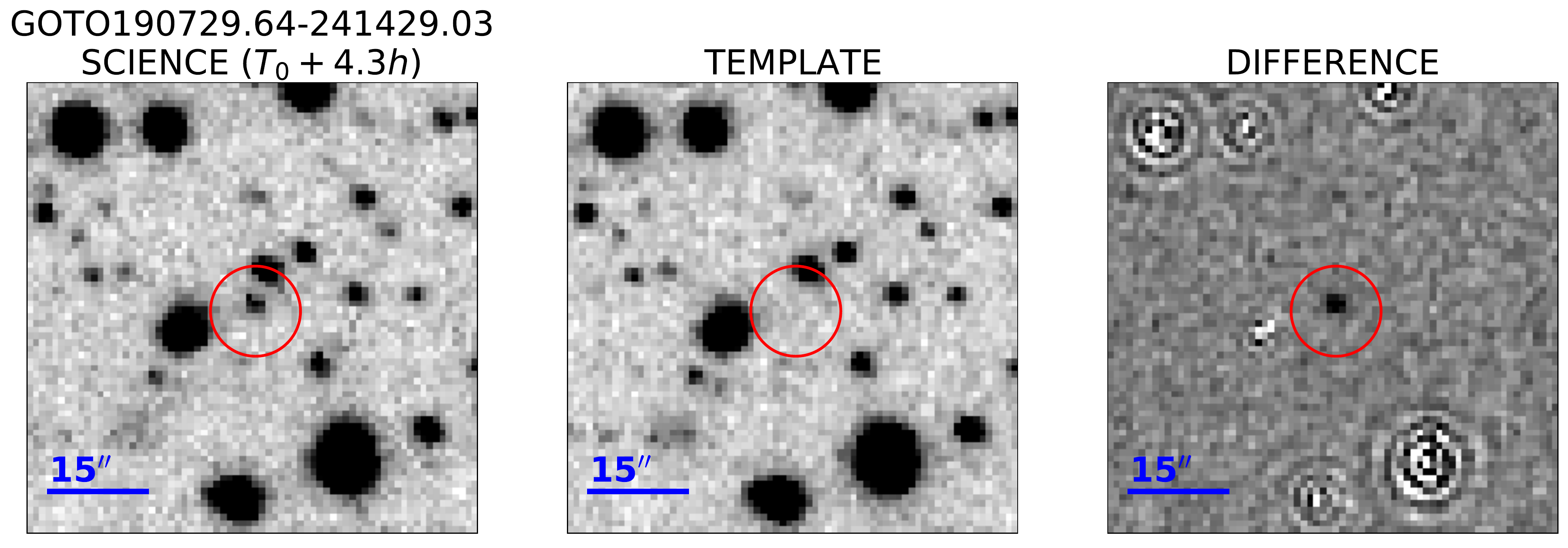}
    \includegraphics[width=\columnwidth]{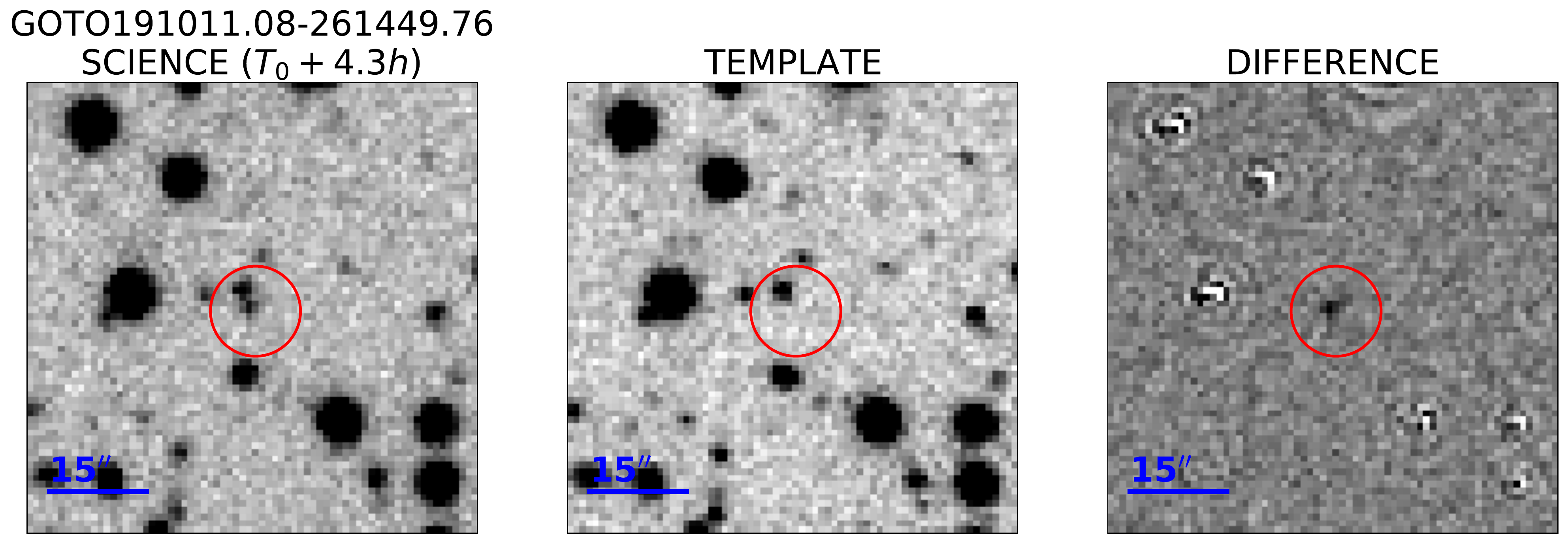}
    \includegraphics[width=\columnwidth]{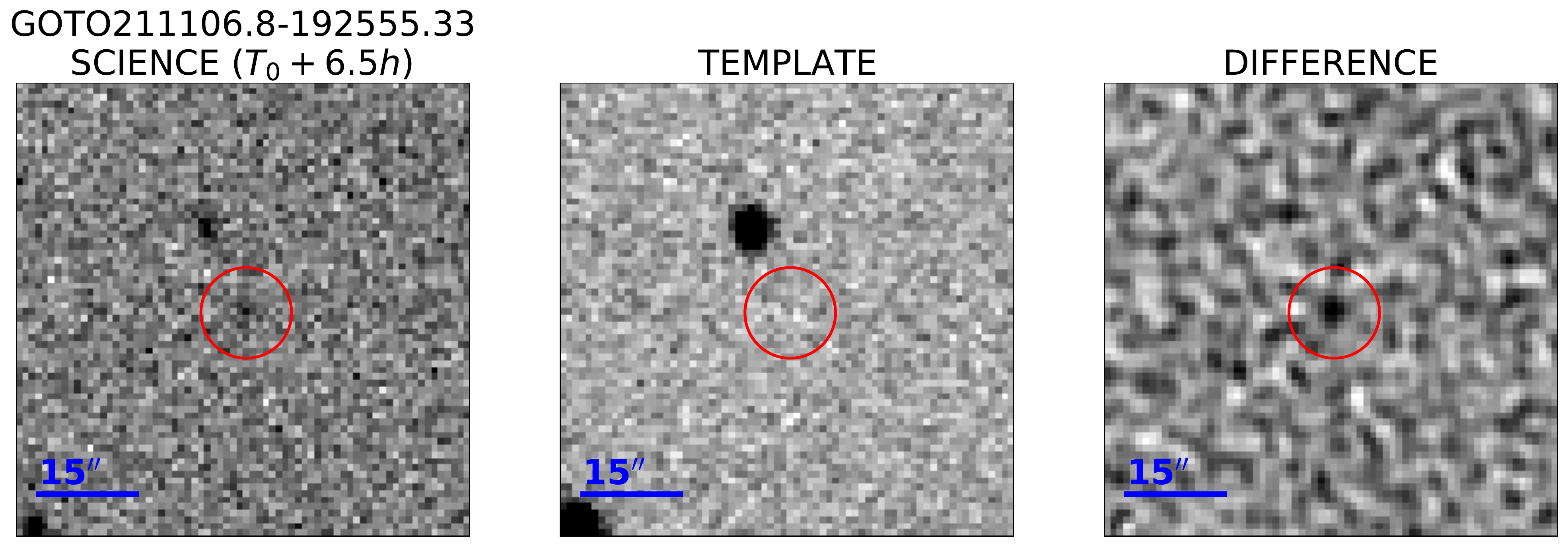}
    \caption{70-by-70 pixel cutout thumbnails of the group 3 candidates. The {\it left}, the {\it middle} and the {\it right} thumbnails are from the science image, the reference image and the difference image, respectively.}\label{fig:cand_thumbnails}
\end{figure}

Group 3 involves 5 candidates: GOTO150228.81-080300.02, GOTO091511.45-003731.96, GOTO190729.64-241429.03, GOTO191011.08-261449.76 and GOTO211106.8-192555.33. Group 3 candidates and group 2 candidates share the same features, except that we do not find any nearby objects on the Pan-STARRS images. This difference makes the group 3 candidates the most interesting as they are unlikely to be stellar objects (see their cutout thumbnails in Figure \ref{fig:cand_thumbnails}). However, there is still a possibility that they are unregistered MPs. To confirm whether they are GRB optical afterglows, we need extra observations to constraint their decay properties.

To conclude, all candidates mentioned in this section could potentially be the GRB optical afterglow, or other transient types. However, we do not have sufficient data to further confirm their origins. 

In our future follow-ups and based on our results, all candidates which pass all of the filtering processes are deemed likely to be associated with the GRB event and deserve extra follow-up. To do this, we need to improve the latency at all stages of our vetting procedures in order to trigger spectroscopy with other facilities nearby. The GRB association can further be confirmed if we obtain a power-law spectrum. Also, extra observations on an identified host galaxy can help to constrain physical properties of the system, such as the star formation rate, metallicity, age, redshift and spatial offsets between the host galaxy and the transient. These observables can help us to identify and improve the confidence of the progenitor origin of the candidate.

\subsection{Expected number of candidates}
The probability of detecting an optical afterglow to a GRB event by GOTO can be estimated by using
\begin{align}
    P_{\rm OA}=P_{\rm cov}\cdot(1-P_{\rm dark})\cdot P_{m<m_{\rm lim}}(t_{\rm obs},\theta_{\rm obs})~.\label{eq:prob_oa}
\end{align}
The probability $P_{\rm cov}$ of covering the GRB position is simply the total coverage of the official \emph{Fermi} healpix skymap. The term $P_{\rm dark}$ is the ``dark'' GRB rate, which is assumed to be the lower limit of $0.4$ in our estimation \citep{gkk11}. $P_{m<m_{\rm lim}}(t_{\rm obs})$ is the probability that the optical afterglow is bright enough to be detected with the an observing angle $\theta_{\rm obs}$ at $t_{\rm obs}$. The observing conditions will affect the image quality and the limiting magnitude of our observations, which could affect the probability of detecting a transient. However, in our probability estimation, we do not take the observing conditions into account.

\begin{figure}
    \centering
    \includegraphics[width=\columnwidth]{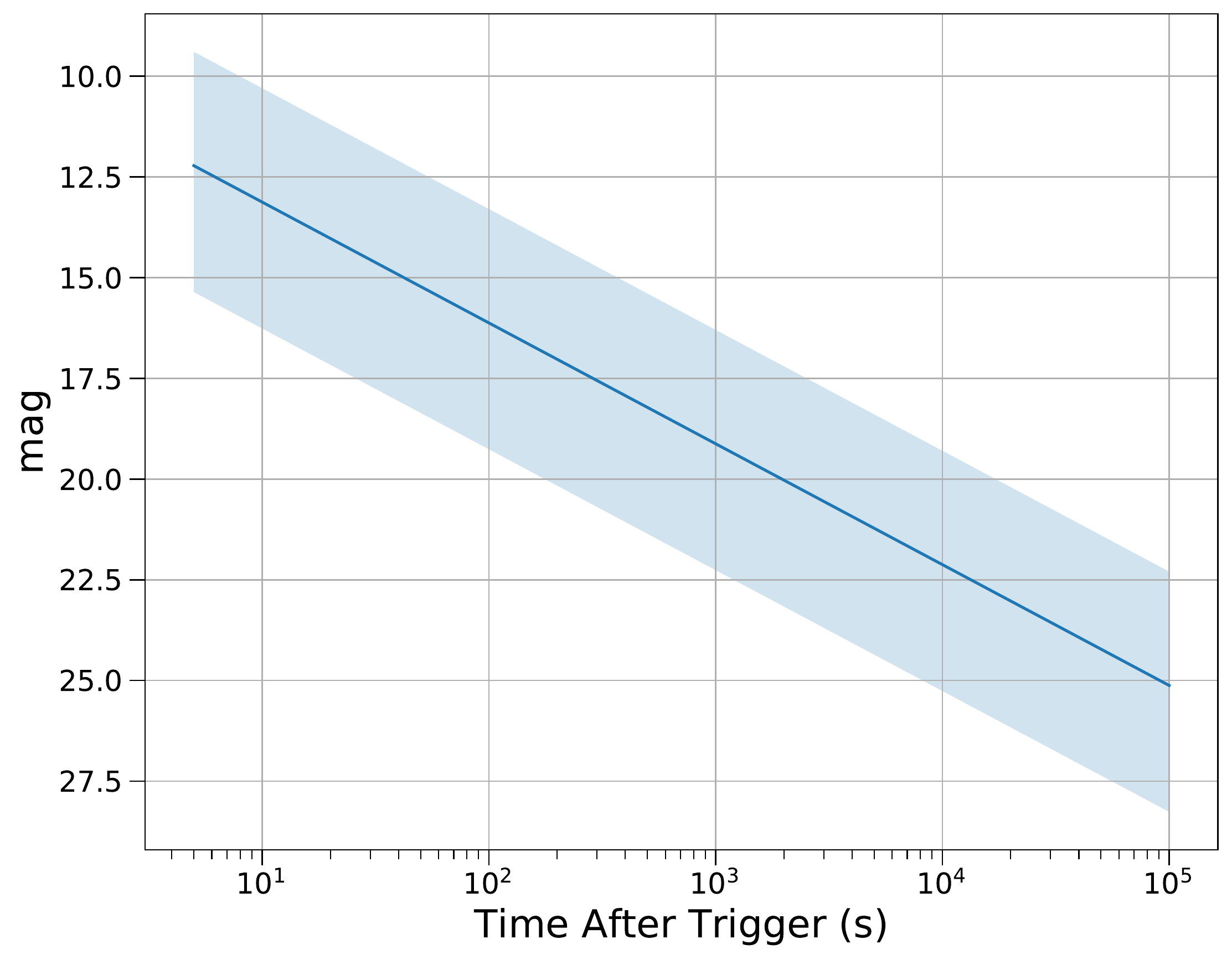}
    \caption{Simulated lightcurve of GRB optical afterglow following power-law decay $t^{-1.2}$. The shaded region indicates the 1-$\sigma$ confidence level.}
    \label{fig:simulated_LC}
\end{figure}

\begin{figure}
    \centering
    \includegraphics[width=\columnwidth]{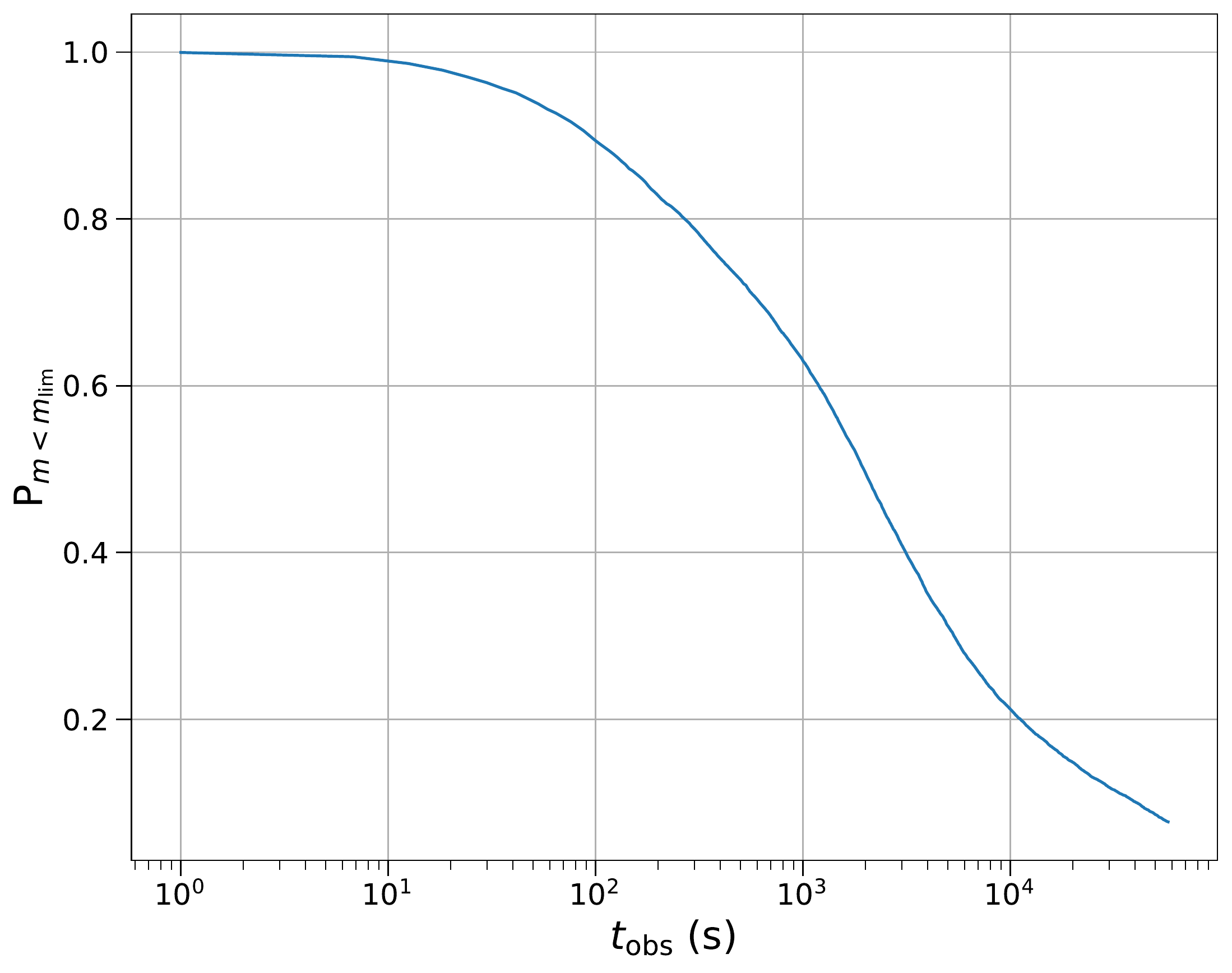}
    \caption{Probability $P_{m<m_{\rm lim}}$ of detecting an optical afterglow brighter than $m=20$ as a function of observing time $t_{\rm obs}$, assuming that the GRB is not dark and that GOTO covers the GRB location at $t_{\rm obs}$.}
    \label{fig:prob_tobs}
\end{figure}

We simulate 14\,302 afterglow lightcurves with a power-law temporal decay $F\propto t^{-1.2}$ using different combination of redshift $z$ and the corrected magnitude $m_0(z=1, t=86\,{\rm s})$. The samples of $z$ and $m_0$ are drawn from the distributions presented in \citet{zga18} and \citet{kkz10}. The simulated lightcurve with 1-$\sigma$ confidence region is shown in Figure \ref{fig:simulated_LC}.
We can obtain $P_{m<m_{\rm lim}}(t_{\rm obs})$ using the simulated lightcurve (see Figure \ref{fig:prob_tobs}). Here, we assume that $\theta_{\rm obs}\approx0$. With the Eq. \ref{eq:prob_oa}, the $P_{\rm OA}$ for each GRB event is shown in Table \ref{tab:events}.

The total expected number of optical afterglows is estimated as the sum of $P_{\rm OA}$ for all events. For our sample of 53 events, we expect to detect $\lesssim3$ afterglows as the upper limit since we do not take the weather conditions into account.


\section{Discussion and Improvements}\label{sec:discussion}
In Eq. \ref{eq:prob_oa}, the factors $P_{\rm dark}$ and $P_{m<m_{\rm lim}}$ depend on the observational properties of the GRB and the limitation of the telescope, which cannot be improved manually. However, the coverage $P_{\rm cov}$ can be improved by optimising the observing strategy and with the addition of more telescopes.

\subsection{Gaussian skymap against official skymap}\label{sec:gaussian_against_official}
The skymaps used for tiling in our current GRB follow-ups are created independently using a two-dimensional normal distribution. The distribution is centered at the GRB position reported by the \emph{Fermi}-GBM with the confidence region of the quadrature sum between the statistical error and the systematic error \citep[3.61 degrees;][]{cbg15}.

The advantage of using manually-created skymaps is to minimize the follow-up response time. The final version of the official skymaps published by \emph{Fermi} have a median delay time of $\sim600\,{\rm s}$. Since the GRB optical afterglow has a fast temporal decay (see the simulated lightcurve in Figure \ref{fig:simulated_LC}), the early post-trigger phase is the essential period for detecting the afterglow emission. If we start follow-up observations after the \emph{Fermi} skymap is released, and assuming that we can cover the correct position within the first 10 minutes, then $\sim30\%$ of the GRBs would become too faint to be detected under GOTO limiting magnitude of 20 (see Figure \ref{fig:prob_tobs}).

\begin{figure}
    \centering
    \includegraphics[width=\columnwidth]{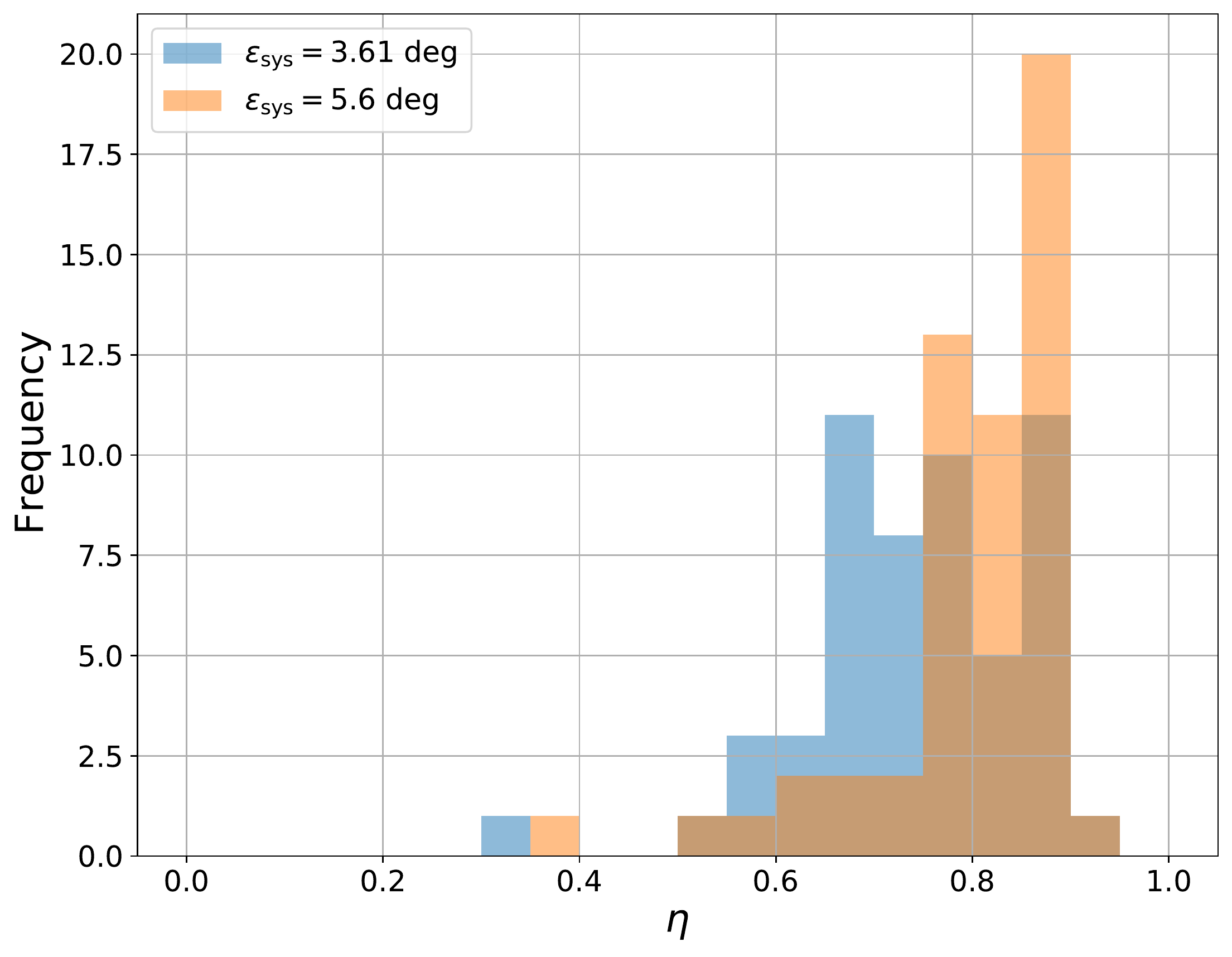}
    \caption{Histogram of the overlapping indices $\eta$ for all 53 \emph{Fermi}-GBM GRB events. The {\it orange} and the {\it blue} histograms represent the skymap created by using the systematic errors of 3.61 and 5.60 degrees, respectively. By optimizing for the systematic error, the median value of $\eta$ has been improved from 0.75 to 0.82.}
    \label{fig:ovl}
\end{figure}

\begin{figure}
    \centering
    \includegraphics[width=\columnwidth]{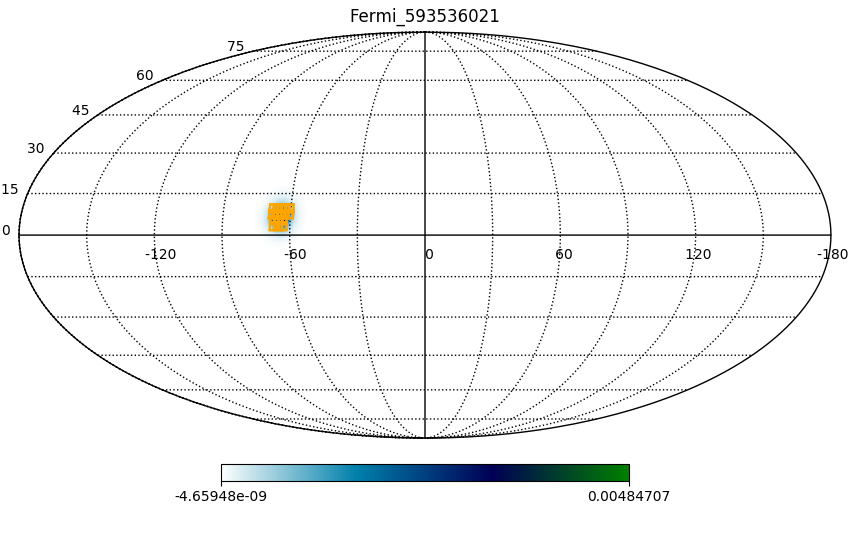}
    \includegraphics[width=\columnwidth]{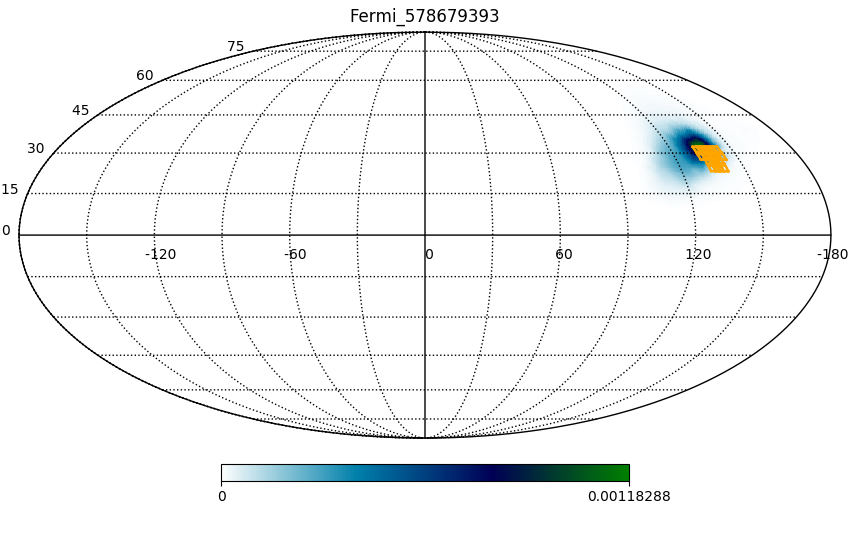}
    \caption{Examples of skymap tiling. The {\it blue} area represents the official skymap and the {\it orange} grids represent the observed tiles generated based on the Gaussian skymap. The {\it top} skymap shows a good tiling case. On the other hand, the observed tiles on the {\it bottom} skymap shows a significant deviation from the center of the skymap.}
    \label{fig:skymap_tiling}
\end{figure}

On the other hand, the official skymap generated by \emph{Fermi}-GBM provides a more precise localization for the GRB detection. Our generated Gaussian skymaps do not always cover the highest confidence region of the official skymap. The similarity of the two skymaps can be measured by using the overlapping index \citep{pmc19}
\begin{align}
    \eta(f,g)=\oint\min[f(\mathbf{r}),g(\mathbf{r})]\,d\Omega~,
\end{align}
where $f$ and $g$ are the probability distributions of the two skymaps. The index $\eta$ closer to 1 implies higher similarity between two skymaps. Figure \ref{fig:ovl} shows that the median value of $\eta$ is 0.75 for all 53 events, which implies that the Gaussian and the official skymaps usually share reasonable amount of overlapping portion. We also show two skymap tiling examples with $\eta=0.87$ (Trigger ID: 593536021) and $\eta=0.53$ (Trigger ID: 578679393) in Figure \ref{fig:skymap_tiling} to illustrate how $\eta$ affects the tiling of the skymap.

\begin{figure}
    \centering
    \includegraphics[width=\columnwidth]{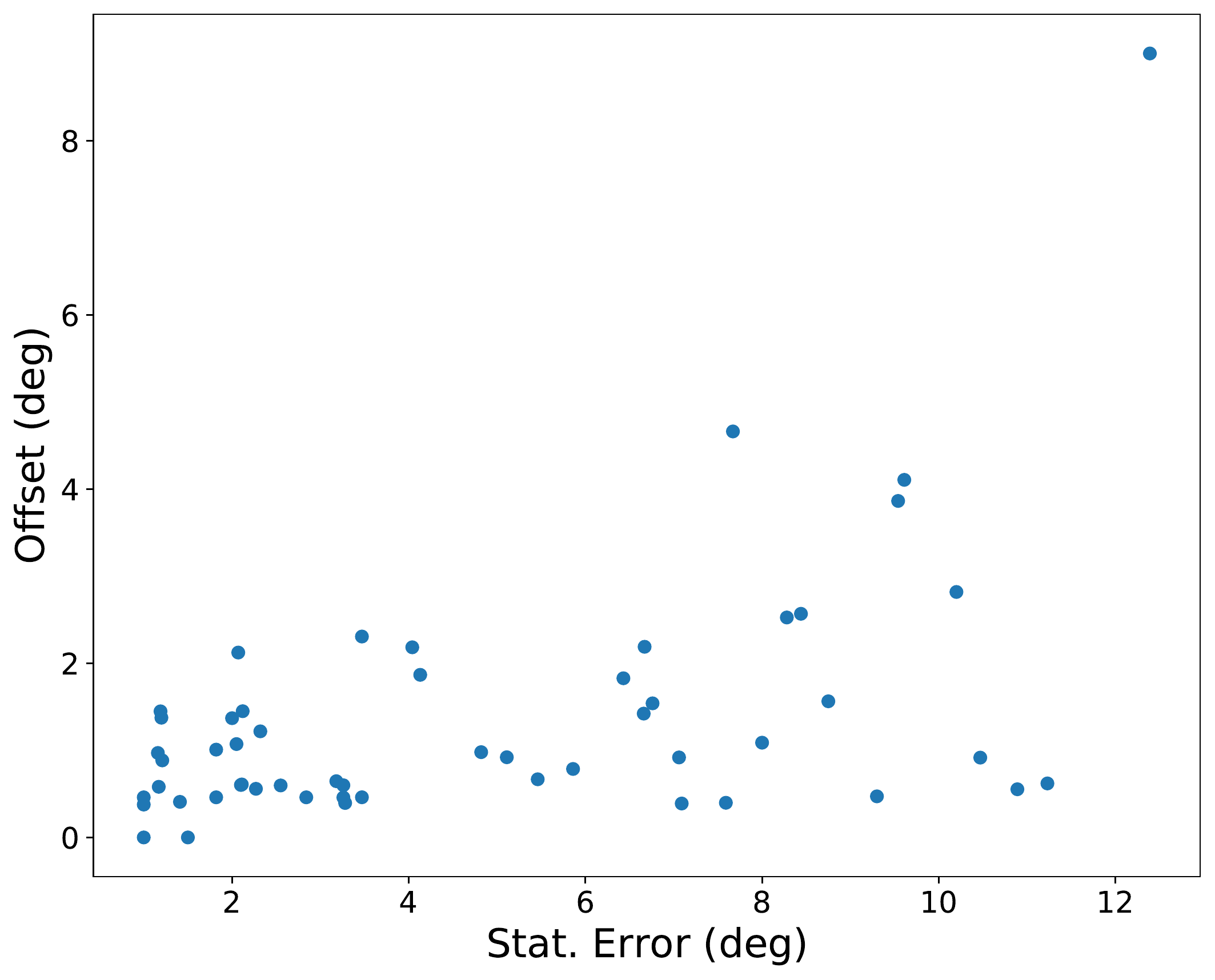}
    \caption{Offset between the peaks of the Gaussian skymap and the official skymap against the \emph{Fermi} reported statistical error. The offsets mostly lie within 3 degrees and shows no correlation with the statistical error.}
    \label{fig:skymap_offset}
\end{figure}

Since we start our observations from the portion of the skymap with the highest probability, we have to confirm that the peaks between our Gaussian skymap and the official skymap do not deviate from each other by more than the size of a single tile (FoV$\approx$20 square degrees). Figure \ref{fig:skymap_offset} shows that most of the offsets are consistent to within 3 degrees. Therefore, the offset between the two skymaps is smaller than the size of a single sky grid.

\begin{figure}
    \centering
    \includegraphics[width=\columnwidth]{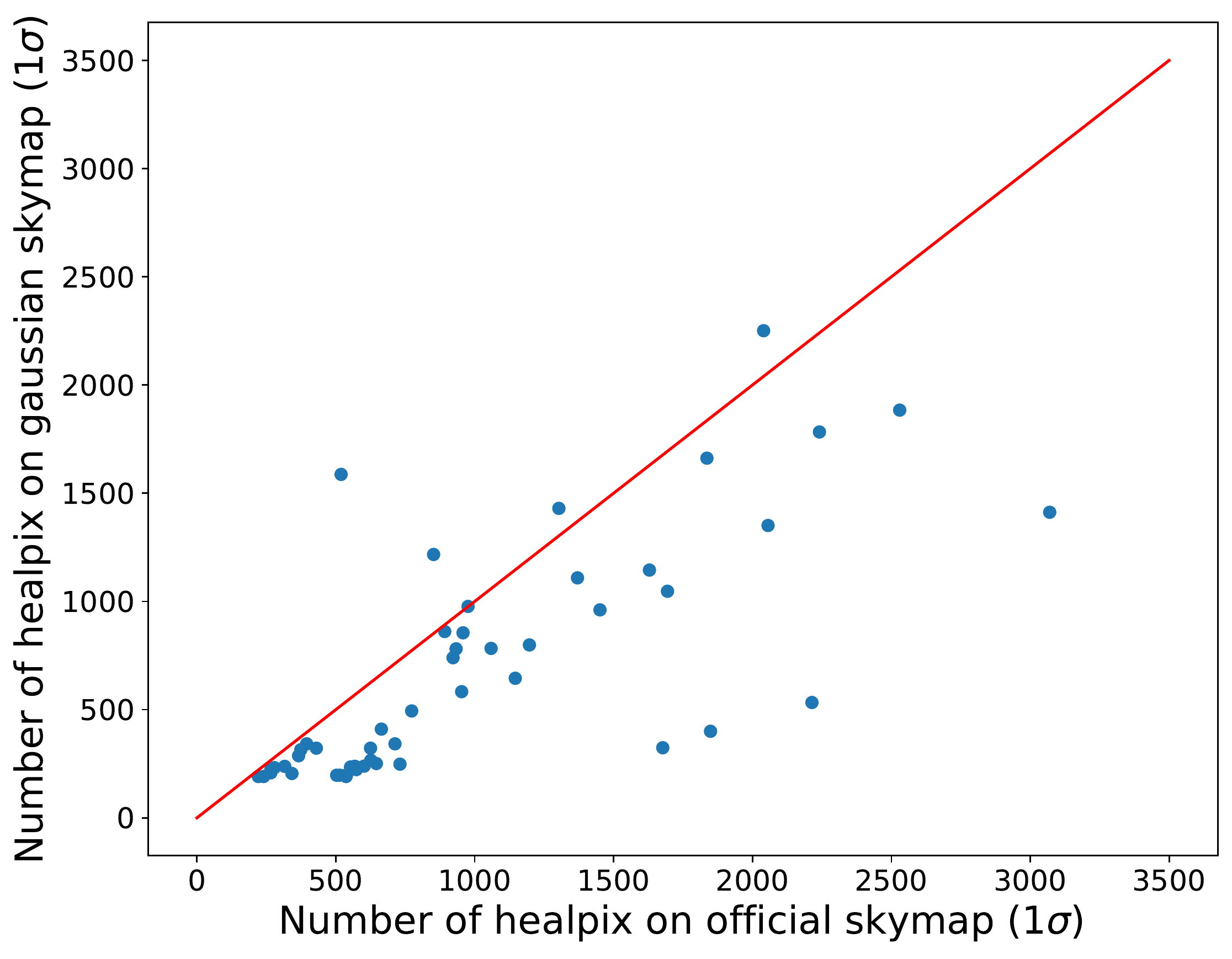}
    \caption{Correlation between the numbers of healpix pixels on the Gaussian skymap (with the systematic error of 3.61 degrees) and the official skymap at $1\sigma$ confidence level. The {\it red} line indicates the number of healpix pixels on both skymaps are equal. Most of the data points lie below the {\it red} line indicating the systematic error is underestimated.}
    \label{fig:num_healpix}
\end{figure}

Other than the offset, the Gaussian skymap should ideally be the same size as the official skymap. If we use the systematic error of $3.61$ degrees to create the intermediary Gaussian skymaps, most of will be smaller in extent than the official ones (see Figure \ref{fig:num_healpix}), which implies that we have underestimated the systematic error. In order to generate a Gaussian skymap with a similar size of the official one, we increase the systematic error to maximize the likelihood function
\begin{align}
    \log\mathcal{L}(\varepsilon_{\rm sys})=-\frac{1}{2}\sum_{i=1}^n\left[\frac{x_i-y_i(\varepsilon_{\rm sys})}{\sigma_i}\right]^2~,
\end{align}
where $x_i$ is the number of pixels on the Healpix grid \citep{ghb05}, which is a pixelization algorithm of 2-sphere with equal area for each individual grid, of order=NESTED as defined by the Fermi skymap and $y_i(\varepsilon_{\rm sys})$ is the number of Healpix pixels on the Gaussian skymap generated with a systematic error $\varepsilon_{\rm sys}$. We apply $\sigma_i=\sqrt{x_i}$ in the likelihood. The optimal value of $\varepsilon_{\rm sys}$ is 5.6 degrees. With the optimal $\varepsilon_{\rm sys}$, the median value of $\eta$ increases from 0.75 to 0.82 (see Figure \ref{fig:ovl}), which indicates that the overall similarity between the Gaussian skymap and the official skymap has been improved.

Despite the high $\eta$, the official skymap cannot be reproduced perfectly by a simple two-dimensional Gaussian distribution. To solve this problem, the manually created Gaussian skymap will be replaced by the official skymap once it is published, and the typical creation time of the official skymap is $\approx10\,{\rm mins}$. The observation schedule will then be updated in response.

\subsection{Tiling strategy}

The size of the official skymaps shows a wide range of variation depending on the uncertainty of the \emph{Fermi}-GBM localization. Even with a high visibility for some events (see Table \ref{tab:events}), the coverages within 16 hours are still low, which indicates that only covering the top five tiles is not enough in our follow-up observations. The median size of the official skymap at $1\sigma$ confidence level is $\approx150$ square degrees. With the GOTO protoype FoV of $20$ square degrees, we need at least 8 tiles to cover 68\% of the skymap.

The solution to this problem is straight forward. Instead of tiling a fixed number of tiles, we should tile a fraction, e.g. 68\%, of the skymap. With this strategy, we can guarantee to cover a reasonable portion of the skymap even with a large uncertainty. However, with this strategy, we require that a minimum fraction, say 68\%, of the skymap can be covered within a certain amount of time. The time constraint should not be set for too long due to the fast-decay nature of the afterglow. According to Figure \ref{fig:prob_tobs}, we estimate that 80\% of the afterglows would fall below the limiting magnitude of GOTO within 2.8 hours.

\subsection{Follow-up Cadence}\label{sec:cadence_improve}
With the current configuration of GOTO, it takes $4\times90$ seconds of exposure per each frame. Based on our 5-tile observing strategy, it takes 30 minutes to complete the whole tiling process. Figure \ref{fig:simulated_LC}
shows that GRB optical afterglow can decay up to $\sim10$ mag within the first 30 minutes after the trigger.

In order to verify the fast-decay nature of GRB optical afterglows, if we can trigger the first follow-up observation within 30 minutes after the trigger, we should start our second follow-up observation immediately afterwards. Since the afterglow decay follows the power-law function, the magnitude should decay much slower on the linear timescale after the first hour of the trigger. To estimate a proper cadence of the following observations, we use
\begin{align}
    \Delta m=-2.5\log\left(\frac{t}{t_0}\right)^{-\alpha}~,
\end{align}
where $\Delta m$ is the magnitude difference between time $t$ and $t_0$. By taking $\Delta m=1$ and $\alpha=1.2$, we obtain $t\approx3.3t_0$, which is our estimate of the cadence for the subsequent observation following the first observation taken at time $t_0$.% if we expect a magnitude change of $\Delta m=1$.

\section{Conclusion}

From February 2019 to June 2020, the GOTO prototype followed-up 93 GRBs detected by \emph{Fermi}-GBM. We conduct an archival search on 53 of the them followed up within 16 hours and without \emph{Swift} joint detections.

We have developed a process to effectively filter detections that are unlikely to be related to a GRB. We divide the process into two parts, automated and manual. The automated process involves real-bogus classification, bad data pre-filtering, catalog cross-matching, minor planet checking and multi-detection filtering. These automated processes narrow down our candidate number from 60\,085 to 6\,276. The manual filtering process involves human vetting, transient cross-matching and forced photometry analysis. In the human vetting step, we finally obtain 119 transient-like objects, 55 of which have nearby galaxies. Transient cross-matching further helps us filter out some GRB-unrelated sources, ultimately resulting in 29 potential candidates. 

We apply forced photometry on those 29 candidates using data obtained by GOTO, ZTF and ATLAS. We analyze their lightcurves and find that 11 of them are unlikely to be related to any GRB. For the rest of the candidates, they can potentially be associated as a GRB afterglow. However, we do not have enough data to solidly constrain their transient types.

We expect to detect $\approx3$ GRB optical afterglows among those 53 events. The estimation is calculated by simulating the lightcurve of the afterglow. The skymap coverage of each event and the dark GRB rate are also considered in the calculation.

In order to improve our performance, we recommend using 5.6 degrees as the systematic error to create a Gaussian skymap for tiling before the official skymap provided by \emph{Fermi} is released. Once the official skymap is released, which should take $\sim10$ minutes after the trigger, we should use this updated skymap for tiling. For the tiling strategy, we should image at least $1\sigma$ of the visible skymap to ensure that a reasonable portion of the skymap is covered. If the first follow-up observation can be triggered within 30 minutes after the GRB alert, we also recommend multiple observations. In order to verify the nature of fast decay, a second follow-up should be performed immediately following the first one.

\section*{Acknowledgements}
We thank the referee, Michael Coughlin for the comments that improved this paper. 
The Gravitational-wave Optical Transient Observer (GOTO) project acknowledges the support of the Monash-Warwick Alliance; Warwick University; Monash University; Sheffield University; University of Leicester; Armagh Observatory \& Planetarium; the National Astronomical Research Institute of Thailand (NARIT); the Instituto de Astrof\'isica de Canarias (IAC) and the University of Turku. RLCS and POB acknowledge support from STFC. RB, MK and DMS acknowledge support from the ERC under the European Union's Horizon 2020 research and innovation programme (grant agreement No. 715051; Spiders).
VSD and MJD acknowledge the support of a Leverhulme Trust Research Project Grant.

\section*{Data availability}
{Data files covering the system throughput and some of the software packages are available via public github repositories under \url{https://github.com/GOTO-OBS/}. Prototype data was mainly used for testing and commissioning and a full release of all data is not foreseen. Some data products will be available as part of planned GOTO public data releases.}

%%%%%%%%%%%%%%%%%%%%%%%%%%%%%%%%%%%%%%%%%%%%%%%%%%


% The all.bib file includes all the required references (and many more)
% Include any additional references there, or put them in-line in the
% text with a reminder to add them

\bibliographystyle{mnras}
\bibliography{all} % if your bibtex file is called example.bib

% \begin{thebibliography}{99}
% \bibitem[\protect\citeauthoryear{Author}{2012}]{Author2012}
% Author A.~N., 2013, Journal of Improbable Astronomy, 1, 1
% \bibitem[\protect\citeauthoryear{Others}{2013}]{Others2013}
% Others S., 2012, Journal of Interesting Stuff, 17, 198
% \end{thebibliography}

%%%%%%%%%%%%%%%%%%%%%%%%%%%%%%%%%%%%%%%%%%%%%%%%%%

%%%%%%%%%%%%%%%%% APPENDICES %%%%%%%%%%%%%%%%%%%%%

% \appendix

% \section{Some extra material}

% If you want to present additional material which would interrupt the flow of the main paper,
% it can be placed in an Appendix which appears after the list of references.

%%%%%%%%%%%%%%%%%%%%%%%%%%%%%%%%%%%%%%%%%%%%%%%%%%

% Don't change these lines
\bsp	% typesetting comment
\label{lastpage}
%Rhaana: I would give the table a more detailed description: say it is GOTO-observed Fermi GBM gamma-ray bursts.
\begin{landscape}
    \begin{table*}
    	\centering
    	\caption{\emph{Fermi}-GRB event table.}\label{tab:events}
    	\label{tab:data_set}
    	\begin{tabular}{lccccccc}
    		\hline
            \textbf{Trigger Number} & \textbf{Event Time} & \textbf{R.A.} & \textbf{Dec.} & \textbf{Error}& \textbf{Response} & \textbf{Skymap Coverage /} & \textbf{Expected}\\ &&&&&\textbf{Time} & \textbf{Visibility}\footnotesize$^a$ & \textbf{Probability}\footnotesize$^b$\\
            &(UTC)&(hh:mm:ss)&(dd:mm:ss)&(Deg.)&(Hrs)&(\%) & (\%)\\
            \hline
            572876510 & 2019-02-26T12:21:46 & 14:57:43.2 & -08:36:36 & 5.11 & 13.911 & 23.6/87.2 & 1.2\\
            573284727 & 2019-03-03T05:45:22 & 19:55:48 & +29:51:00 & 23.97 & 0.476 & 22.7/60.7 & 7.2\\
            573604668 & 2019-03-06T22:37:43 & 15:24:16.8 & -00:22:48 & 2.55 & 0.0392 & 87.3/100 & 45.2\\
            574345067 & 2019-03-15T12:17:42 & 09:46:14.4 & -11:09:00	& 2.11 & 8.146 & 71.8/100 & 5.2\\
            574676902 & 2019-03-19T08:28:18 & 12:50:24 & -04:48:36 & 3.26 & 14.373 & 100/100 & 5.0\\
            575018216 & 2019-03-23T07:16:52 & 09:28:16.8 & +02:50:24 & 2.07 & 13.478 & 60.9/100 & 3.2\\
            576241792 & 2019-04-06T11:09:47 & 19:05:21.6 & +61:30:00 & 7.09 & 14.557 & 32.5/96.5 & 1.6\\
            576265958 & 2019-04-06T17:52:33 & 19:30:33.6 & +26:47:24 & 5.46 & 9.077 & 19.3/100 & 1.3\\
            578252995 & 2019-04-29T17:49:51 & 13:20:12 & -08:00:00 & 8.00 & 3.121 & 3.7/99.8 & 0.4\\
            578679393 & 2019-05-04T16:16:28 & 09:13:57.6 & +28:41:24 & 9.61 & 5.045 & 17.9/99.8 & 1.7\\
            578711654 & 2019-05-05T01:14:09 & 22:21:33.6 & +42:10:48 & 9.54 & 2.391 & 10.5/82.8 & 1.4\\
            578903308 & 2019-05-07T06:28:23 & 10:23:50.4 & -12:48:00 & 4.82 & 14.583 & 53.0/99.6 & 2.6\\
            578963794 & 2019-05-07T23:16:30 & 19:11:16.8 & -22:49:12 & 1.19 & 3.567 & 54.2/93.8 & 5.9\\
            579036175 & 2019-05-08T19:22:50 & 11:51:50.4 & +23:31:48 & 2.12 & 1.688 & 17.1/100 & 2.8\\
            579814215 & 2019-05-17T19:30:10 & 18:00:04.8 & +25:46:12 & 1.20 & 3.250 & 77.7/100 & 9.1\\
            580437952 & 2019-05-25T00:45:48 & 22:32:04.8 & +05:27:00 & 4.04 & 2.712 & 12.7/86.5 & 1.6\\
            580904353 & 2019-05-30T10:19:08 & 08:03:02.4 & +35:30:00 & 10.20 & 11.028 & 25.3/30.8 & 1.5\\
            581068049 & 2019-06-01T07:47:24 & 10:51:55.2 & +54:35:24 & 8.28 & 13.581 & 25.3/99.8 & 1.3\\
            581281470 & 2019-06-03T19:04:26 & 01:20:19.2 & +40:54:36 & 5.86 & 9.579 & 13.0/30.8 & 0.9\\
            581337762 & 2019-06-04T10:42:37 & 22:50:12 & +46:22:12 & 1.00 & 14.974 & 51.5/54.5 & 2.5\\
            581469752 & 2019-06-05T23:22:27 & 22:28:57.6 & +04:47:24 & 6.76 & 3.691 & 9.5/100 & 1.0\\
            581882394 & 2019-06-10T17:59:50 & 21:49:31.2 & +42:25:12 & 1.21 & 6.603 & 50.2/100 & 4.0\\
            581889628 & 2019-06-10T20:00:24 & 20:59:19.2 & -15:55:48 & 11.23 & 5.798 & 8.1/92.7 & 0.7\\
            582004649 & 2019-06-12T03:57:25 & 14:55:48 & +62:06:00 & 9.30 & 0.190 & 10.1/61.3 & 4.1\\
            582304592 & 2019-06-15T15:16:27 & 12:45:36 & +49:22:48 & 2.32 & 6.748 & 47.0/100 & 3.7\\
            582596766 & 2019-06-19T00:26:02 & 23:17:14.4 & +12:51:36 & 2.05 & 1.916 & 50.4/100 & 7.8\\
            582725415 & 2019-06-20T12:10:11 & 10:48:19.2 & +30:28:48 & 1.16 & 9.348 & 100/100 & 6.7\\
            584590606 & 2019-07-12T02:16:42 & 19:13:33.6 & +56:09:00 & 7.59 & 0.186 & 13.3/100 & 5.5\\
            585007213 & 2019-07-16T22:00:08 & 23:02:31.2 & -00:49:48 & 6.43 & 2.811 & 0.02/100 & 0\\
            585559462 & 2019-07-23T07:24:18 & 19:17:52.8 & +25:13:12 & 10.47 & 13.988 & 5.9/100 & 0.3\\
            585847498 & 2019-07-26T15:24:54 & 20:41:02.4 & +34:17:24 & 1.17 & 5.944 & 100/100 & 8.5\\
            592297741 & 2019-10-09T07:08:57 & 01:47:14.4 & +65:43:48 & 1.41 & 12.773 & 97.6/100 & 5.4\\
            593045905 & 2019-10-17T22:58:21 & 08:58:12 & +15:19:48 & 12.39 & 5.196 & 0.7/99.8 & 0.1 \\
            593176520 & 2019-10-19T11:15:15 & 07:18:14.4 & +62:05:24 & 3.28 & 13.831 & 19.6/93.2 & 1.0\\
            593419810 & 2019-10-22T06:50:05 & 18:12:33.6 & -23:06:00 & 3.18 & 12.870 & 21.1/22.1 & 1.1\\
            593536021 & 2019-10-23T15:06:57 & 19:41:02.4 & +06:10:48 & 1.82 & 4.573 & 73.0/100 & 7.2\\
            593928606 & 2019-10-28T04:10:01 & 21:16:48 & -11:17:24 & 8.75 & 15.459 & 20.0/99.6 & 0.9\\
            593964489 & 2019-10-28T14:08:04 & 18:27:31.2 & +69:59:24 & 7.67 & 6.0339 & 49.2/97.8 & 4.2\\
            596387570 & 2019-11-25T15:12:46 & 23:34:09.6 & +18:12:00 & 4.13 & 4.207 & 38.3/100 & 3.9\\
            596786686 & 2019-11-30T06:04:41 & 23:17:36 & +63:04:48 & 2.27 & 14.151 & 35.4/99.9 & 1.8\\
            \hline
    	\end{tabular}
    \end{table*}
\end{landscape}
\begin{landscape}
    \begin{table*}
    	\centering
    	\begin{tabular}{lccccccc}
    		\hline
            \textbf{Trigger Number} & \textbf{Event Time} & \textbf{R.A.} & \textbf{Dec.} & \textbf{Error}& \textbf{Response} & \textbf{Skymap Coverage /} & \textbf{Expected}\\ &&&&&\textbf{Time} & \textbf{Visibility}\footnotesize$^a$ & \textbf{Probability}\footnotesize$^b$\\
            &(UTC)&(Deg.)&(Deg.)&(Deg.)&(Hrs)&(\%) & (\%)\\
            \hline
            597955752 & 2019-12-13T18:49:08 & 22:04:14.4 & -13:56:24 & 6.67 & 0.665 & 36.0/94.3 & 14.0\\
            598951521 & 2019-12-25T07:25:17 & 06:21:57.6 & -17:21:00 & 6.66 & 15.246 & 35.7/82.3 & 2.2\\
            598988276 & 2019-12-25T17:37:52 & 09:43:12 & -07:10:48 & 2.84 & 7.410 & 100/100 & 7.6\\
            600448273 & 2020-01-11T15:11:08 & 06:57:57.6 & +31:43:12 & 3.47 & 11.244 & 88.6/100 & 5.9\\
            600525396 & 2020-01-12T12:36:31 & 10:00:31.2 & +64:24:36 & 2.00 & 9.431 & 90.3/100 & 6.5\\
            601677816 & 2020-01-25T20:43:31 & 00:29:48 & +64:41:24 & 1.00 & 0.202 & 100/100 & 40.6\\
            601841483 & 2020-01-27T18:11:19 & 05:03:33.6 & +20:04:12 & 3.26 & 1.740 & 45.5/100 & 11.8\\
            603142206 & 2020-02-11T19:30:01 & 07:24:33.6 & +59:00:36 & 10.89 & 0.741 & 17.1/99.8 & 6.4\\
            603849435 & 2020-02-19T23:57:10 & 17:37:55.2 & +08:23:24 & 1.00 & 4.818 & 87.5/100 & 9.6\\
            610450873 & 2020-05-06T09:41:09 & 12:43:40.8 & +40:14:24 & 3.47 & 12.422 & 7.2/100 & 0.5\\
            610800081 & 2020-05-10T10:41:17 & 10:20:19.2 & -01:55:48 & 1.50 & 10.659 & 63.7/100 & 5.8\\
            611434353 & 2020-05-17T18:52:28 & 07:40:48 & +29:25:48 & 7.06 & 2.329 & 6.1/95.3 & 2.0\\
            613212114 & 2020-06-07T08:41:50 & 11:26:00 & +30:52:48 & 2.10 & 12.744 & 67.0/100 & 4.1\\
    		\hline
    		\multicolumn{8}{l}{\footnotesize$^a$ Calculated with the time constraints of 16 hours after the trigger. Official \emph{Fermi} skymap is used.}\\
            \multicolumn{8}{l}{\footnotesize$^b$ The probability of detecting the GRB optical afterglow estimated base on Eq. \ref{eq:prob_oa}.}\\
    	\end{tabular}
    \end{table*}
\end{landscape}
\end{document}